\documentclass[conference]{IEEEtran}
\IEEEoverridecommandlockouts
\usepackage{cite}
\usepackage{amsmath,amssymb,amsfonts}
\usepackage{algorithmic}
\usepackage{graphicx}
\usepackage{textcomp}
\usepackage{xcolor}
\def\BibTeX{{\rm B\kern-.05em{\sc i\kern-.025em b}\kern-.08em
    T\kern-.1667em\lower.7ex\hbox{E}\kern-.125emX}}
    
\usepackage{algorithm}
\usepackage{booktabs}
\usepackage{multirow}
\usepackage{diagbox}
\usepackage{enumitem}

\begin{document}

\title{Influential Recommender System\thanks{Pengfei Zhao is the corresponding author.}}


\author{
\IEEEauthorblockN{Haoren Zhu\IEEEauthorrefmark{1}, Hao Ge\IEEEauthorrefmark{1}, Xiaodong Gu\IEEEauthorrefmark{2}, Pengfei Zhao\IEEEauthorrefmark{3}, Dik Lun Lee\IEEEauthorrefmark{1}}
\IEEEauthorblockA{\IEEEauthorrefmark{1}Hong Kong University of Science and Technology, \emph{\{hzhual, hge, dlee\}@cse.ust.hk}}
\IEEEauthorblockA{\IEEEauthorrefmark{2}Shanghai Jiao Tong University, \emph{xiaodong.gu@sjtu.edu.cn}}
\IEEEauthorblockA{\IEEEauthorrefmark{3}BNU-HKBU United International College, \emph{ericpfzhao@uic.edu.cn}}
}

\maketitle

\begin{abstract}
Traditional recommender systems are typically passive in that they try to adapt their recommendations to the user's historical interests. However, it is highly desirable for commercial applications, such as e-commerce, advertisement placement, and news portals, to be able to expand the users' interests so that they would accept items that they were not originally aware of or interested in to increase customer interactions. In this paper, we present Influential Recommender System (IRS), a new recommendation paradigm that aims to proactively lead a user to like a given objective item by progressively recommending to the user a sequence of carefully selected items (called an influence path). We propose the Influential Recommender Network (IRN), which is a Transformer-based sequential model to encode the items' sequential dependencies. Since different people react to external influences differently, we introduce the Personalized Impressionability Mask (PIM) to model how receptive a user is to external influence to generate the most effective influence path for the user. To evaluate IRN, we design several performance metrics to measure whether or not the influence path can smoothly expand the user interest to include the objective item while maintaining the user's satisfaction with the recommendation. Experimental results show that IRN significantly outperforms the baseline recommenders and demonstrates its capability of influencing users' interests.
\end{abstract}


\section{Introduction}
\label{sec:Introduction}
\textit{Recommender system} (\textit{RS}) has created huge business values for applications such as e-commerce, advertising, and streaming media. Traditional \textit{RS} tries to learn users' internal interests from their historical user-item interactions and recommend items judged to be relevant to the users' interests. 
Despite the recent advancements of \textit{RS} in modeling the users' long/short-term interests, most existing works on \textit{RS}s are passive in that their recommendations are driven by users' historical interests. If the user does not respond to a recommendation positively, these \textit{RS}s would try to please the user by selecting another item that better fits the user's interest. In other words, they model the users' interests and fit their recommendations to the user's internal interests instead of proactively developing their external interests. We refer to this \textit{RS} approach as \textit{User-oriented Recommender System} (\textit{URS}).

For many commercial applications, a business may want its recommendation system to be able to \textit{influence} users to like an objective item that they originally may not be interested in by expanding the users' current interests. This allows a business to develop effective branding and marketing strategies by enticing users to pay attention to some promotional or new-arrival items. A \textit{URS} would not be able to achieve this goal since it only displays items within the users' interests. On the other hand, a straightforward advertisement placement system will simply display the objective items to the users, amounting to hard-selling the items to the users. Neither is desirable from a business or user point of view.
In this paper, we propose a new \textit{RS} paradigm named \textit{Influential Recommender System} (\textit{IRS}). \textit{IRS} aims to influence a user $u$'s interest towards a given objective item $i_t$ by recommending a sequential list of items (called \textit{influence path}), which not only follows $u$'s historical interests but also progressively broadens $u$'s interests towards the objective $i_t$. 

\begin{figure}[tb]
    \flushleft
	\includegraphics[width=0.5\textwidth]{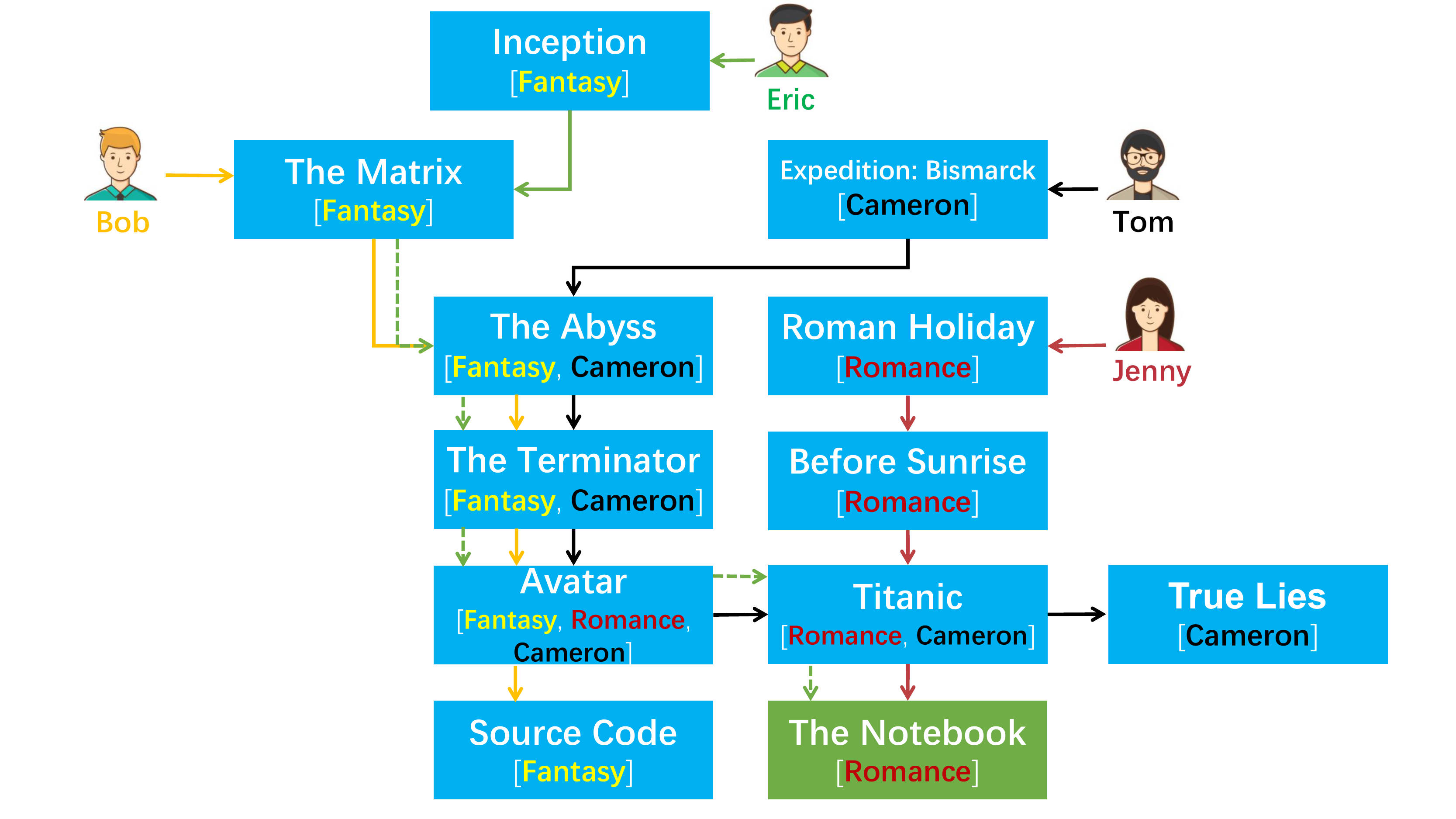}
	\caption{Illustration of Influential Recommender System.}
	\label{fig:example}
\end{figure}

Figure~\ref{fig:example} illustrates the idea of \textit{IRS} in a movie recommendation scenario. Solid arrows indicate users' viewing histories (one color per user), which reflect the users' chronological behavioral sequences of movie watching. These sequences reflect not only the users' interests profile but also the movies’ sequential dependencies. In the example, Bob is interested in fantasy movies, with a watch history of \textit{The Matrix} $\rightarrow$ \textit{The Abyss} $\rightarrow$ \textit{The Terminator} $\rightarrow$ \textit{Avatar} $\rightarrow$ \textit{Source Code},
in which all movies are of the fantasy genre. Similarly, Tom is a fan of director James Cameron, and Jenny likes romance movies.

Assume that the goal of \textit{IRS} is to influence Eric to watch the objective movie \textit{The Notebook}, a classic romance movie. From Eric's watch history, \textit{Inception} $\rightarrow$ \textit{The Matrix}, it is fair to infer that Eric has a high chance of accepting fantasy movies, but how can \textit{IRS} lead him to watch the romance movie \textit{The Notebook}?

We notice that Eric and Tom both watched \textit{The Matrix}. Thus, there is a good chance for Eric to accept Tom's behavioral sequence, i.e., 
\textit{The Abyss} $\rightarrow$ \textit{The Terminator} $\rightarrow$ \textit{Avatar}.
After Eric accepts \textit{Avatar}, a romance movie directed by Cameron, we are more confident to say that Eric has developed a certain level of interest in Cameron and the romance genre. Taking one step forward, we can recommend \textit{Titanic}, another classic romance movie directed by Cameron, to Eric. If Eric accepts it, his interest in romance movies is further strengthened. Following Jenny's sequence, recommending \textit{The Notebook} as the next movie for Eric becomes natural and persuasive. This potential recommendation sequence (called \textit{influence path}) is indicated by the green dash arrows. This example illustrates the goal of \textit{IRS}, that is, the progressive expansion of a user's interest towards the objective item along an enticing, smooth influence path learned from the large historical user data.

There are key differences between \textit{URS} and \textit{IRS}. For \textit{URS}, users play the central role, and \textit{URS} aims to adapt to the users' historical interests, while for \textit{IRS}, the platform takes the proactive role to leading the users to like the preset objective items by progressively recommending a carefully selected sequence of items. Filling the gap from \textit{URS} to \textit{IRS} presents several challenges. 
First, the influence path should (i) not deviate too much from the user’s historical interest in order to maintain the user’s trust and satisfaction with the system and (ii) progressively recommend items related to the objective item so that the user’s interest in the objective item can be developed and the transition from the user's interest towards the objective item is gradual and smooth. 
Second, user-item interactions have sequential patterns, i.e., the next item/action depends more on items/actions the user engaged in recently \cite{tang2018personalized}. Thus, the generation of an influence path must align with the sequential dependencies among the items. 
Third, a user's preference for external influence is personal. That is, some users may be extrovert, i.e., they favor exploration and thus can be easily affected by external influences, but some may favor staying in their comfort zone and are difficult to be persuaded to accept new interests. For the former, influence paths could be more aggressive, i.e., they can deviate more from the users' existing interests to reach the objective item in fewer steps; for the latter, the reverse should be taken. We introduce the notion of \textit{Impressionability} to account for the users' different preferences of external influence.

To meet these challenges, we propose \textit{Influential Recommender Network} (\textit{IRN}), which is adapted based on the widely used \textit{Transformer} model \cite{vaswani2017attention} in natural language processing (\textit{NLP}).
Both the user-item interaction sequences and \textit{NLP} sentences have a sequential structure, and the dependencies among items/words exist not only in adjacent items/words but also among distant items/words, which can be well captured by the multi-head attention mechanism. 
Furthermore, to accommodate the influence of the objective item and to personalize the generation of the influence path for individual users with different degrees of impressionability, we design a new masking scheme termed \textit{Personalized Impressionability Mask} (\textit{PIM}) that incorporates user embedding into the calculation of item attention weights. 

To the best of our knowledge, we are the first to propose influential recommendations. Consequently, there are no standard performance metrics or similar approaches to serve as baselines for performance evaluation. 
To overcome these problems, we design new performance metrics that can be used to measure whether the proposed \textit{IRS} frameworks can lead the user to the objective item while satisfying the user's current interest. As explained in Section~\ref{sec:experiment}, we need to determine how much a user is interested in an unseen item, which cannot be obtained from offline datasets. Thus, we propose to use an \textit{Evaluator} to simulate the reaction of a user to an unseen item. 


The contribution of this paper is multi-fold:
\begin{enumerate}
\item We propose a new paradigm of recommendation named \textit{Influential Recommender System} (\textit{IRS}), which proactively leads the user towards a given objective item instead of passively adapting the recommendations to the user’s historical interests. To the best of our knowledge, \textit{IRS} is the first to infuse proactivity and influence to \textit{RS}s.

\item We design \textit{Influential Recommender Network} (\textit{IRN}), a Transformer-based model to encode the items' sequential dependencies. In order to model the user's personalized preference of external influence, we extend Transformer by adding user embedding to the attention mechanism.


\item We design several performance metrics and offline experiments specifically for evaluating \textit{IRS}. Experimental results show that our proposed \textit{IRN} significantly outperforms the baseline recommenders and can smoothly and progressively influence a user's interest.

\end{enumerate}

The remaining parts of the paper are organized as follows.
We briefly review the related work 
in Section \ref{sec:relatedwork}. 
In Section \ref{sec:framework}, we formulate the problem of \textit{IRS} and propose three frameworks, including two solutions adapted from existing \textit{URS} and our \textit{Influential Recommender Network}. 
In Section \ref{sec:experiment}, we first introduce the experimental setup and \textit{IRS} evaluation and then present a detailed discussion of the experimental results. 
Finally, we conclude our work and summarize the future directions in Section \ref{sec:conclusion}.

\section{Related Work}
\label{sec:relatedwork}
\subsection{Bayesian Persuasion}
There have been decades of studies on how an entity (e.g., person, group, etc.) can be persuaded to change its action \cite{chaiken1987heuristic, kamenica2011bayesian, ajzen2012attitudes, jadbabaie2012non}. Persuasion can be defined as influencing behavior via the provision of information. \textit{Bayesian Persuasion} studies the persuasion problem in a rational (Bayesian) manner \cite{kamenica2011bayesian, bergemann2016information, kamenica2019bayesian}. The problem can be abstracted as a sender designing a signaling plan with the right amount of information disclosed so as to influence the behaviors of another receiver. Later, extensive works have applied the information design of \textit{Bayesian Persuasion} to real-world scenarios, such as social learning for online recommendation platform \cite{che2018recommender}, products advertisement \cite{arieli2019private}, and matching markets \cite{romanyuk2019cream}, etc. In this paper, we consider the persuasion problem from the perspective of \textit{RS}, which aims at changing the user to like an objective item by recommending a selected sequence of items to the user. In contrast to the existing works that model the persuasion problem as an information design and mechanism design problem requiring domain knowledge, our proposed work is data-driven where the persuasion is done by designing an influence path with proper machine learning techniques. To the best of our knowledge, we are the first to integrate \textit{RS} with persuasion behavior. 

\subsection{Sequential Recommendation System}
\label{subsec:seq_rs}
Given the natural sequential order of user-item interactions, the sequential recommender system (\textit{SRS}) tries to incorporate sequential signals into the next-item recommendation task. Due to the diversity and complexity of user-item interactions, \textit{SRS}s have to face several challenges \cite{ShoujingWang-ijcai2019}. 
(1) Learning higher-order long/short-term sequential dependencies. Compared to lower-order sequential, which are relatively simple and can be easily modeled with Markov chain or factorization machines, long/short-term sequential dependencies are much more complex and harder to capture because of their complicated multi-level cascading dependencies crossing multiple user-item interactions. Various approaches such as recurrent neural networks (\textit{RNN}) based models \cite{Wu2017RRN, hidasi2015session, bharadhwaj2018recgan}, graph-based models \cite{xiang2010temporal, wu2019session, wang2020global} have been applied to \textit{SRS}s due to their strength in modeling long/short-term interests.
(2) Handling user-item interaction sequences with flexible order and noise. \cite{tang2018personalized,Yuan2019} exploits the property of convolutional neural networks (\textit{CNN}) to model the flexible orders of user-item interactions. Attention models have been employed to emphasize relevant and important interactions in a sequence while downplaying irrelevant ones \cite{kang2018self, wang2018attentionbased, Ying2018, sun2019bert4rec}. (3) Handling user-item interaction sequences with heterogeneous relations and hierarchical structures. \cite{Tang2019, Wang2019} proposed mixture models to tackle the problem. The difference between \textit{SRS} and \textit{IRS} is, \textit{SRS} models the items' sequential dependencies for the next-item recommendation task, which belongs to the \textit{URS} category, while \textit{IRS} utilizes the sequential dependency information to generate influence paths towards the objective item.

\section{Influential Recommender System}
\label{sec:framework}

\subsection{Problem Definition}
In sequential recommender systems, user-item interactions are represented as sequences. For a given dataset, we denote the set of all user-item interaction sequences as $S=\{s_1,...,s_{|S|}\}$, which has size $|S|$. Each sequence $s=\{i_1,...,i_{m}\}$ consists of an ordered list of $m$ items. All the items occurring in $S$ constitute the item set $I=\{i_1,...,i_{|I|}\}$ of size $|I|$. 
Let $U$ denote the set of all users, and $S_u$ denote the set of sequences belonging to $u$. We have $S = \mathop{\bigcup}_{u=1}^{|U|} S_u$, where $|U|$ is the number of users. 

\begin{figure}[tb]
	\centering
	\includegraphics[width=0.47\textwidth]{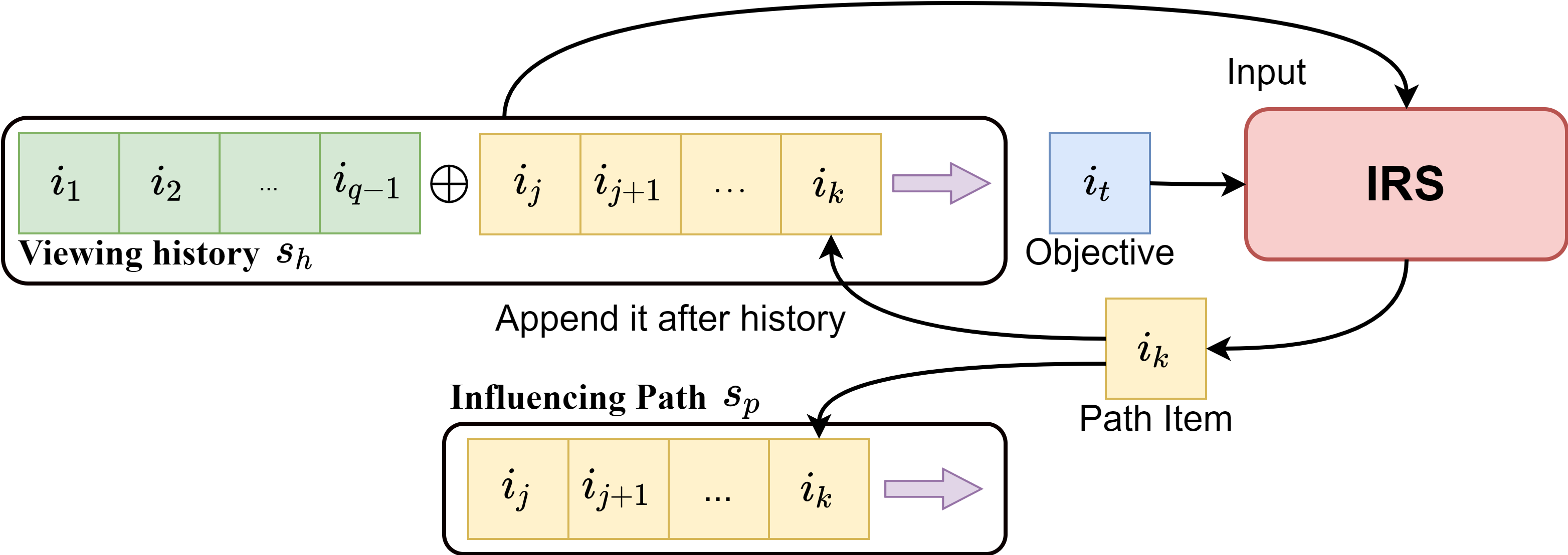}
	\caption{Illustration of the Influential Recommendation Process. \textit{IRS} recursively recommends a path item $i_k$ until reaching the objective item $i_t$ or exceeding the maximum length allowed. All the path items constitute the influence path $s_p$. }
	\label{fig:eval_setting}
\end{figure}

The problem of influential recommendation is defined as:
given a user $u$'s viewing history $s_h$ and a selected objective item $i_{t}$, produce a sequence of items, named \textit{influence path} (denoted as $s_p$), which starts from $u$'s original interests and leads $u$ towards $i_{t}$ progressively and smoothly. 
Figure~\ref{fig:eval_setting} illustrates the idea.
An influence path aims to meet two criteria: a path item $i_k$ should expand the user's interests toward the objective $i_t$ and simultaneously satisfy the user's current interests so that the user will not lose patience to the recommendations. The smoothness of the path is also critical since an abrupt recommendation may upset the user. 
Note that $i_t$ can be picked according to the needs of the specific application. 
In this work, we use random $i_t$ to avoid the bias incurred by manual selection. 

For simplicity, we assume that the user will passively accept any recommendation, and we employ probability measures to evaluate the user's satisfaction with the influence path. Algorithm \ref{al:ppath} demonstrates the overall workflow of \textit{IRS}. 
The main goal of \textit{IRS} is to learn the recommender function $\mathcal{F}$ that progressively generates the influence path $s_p$ to lead the user towards the objective item and simultaneously satisfy the user's interests. To achieve this goal, we present two baseline solutions adapted from existing \textit{URS} and then our novel influential recommender model.  

\begin{algorithm}
\caption{Generate influence path $s_p$ for user $u$}\label{al:ppath}
\begin{algorithmic}[1]
\STATE \textbf{Input}: Interaction history $s_h$, the objective item $i_{t}$, the maximum path length $M$
\STATE \textbf{Output}: influence path $s_p$
\STATE $s_p$ = [ ]
\WHILE {$|s_p| \leq M$}
  \STATE $i = \mathcal{F}(s_h, i_{t}, s_p)$
  \STATE $s_p = s_p + i$
  \IF {$i$ is $i_t$}
    \STATE break
  \ENDIF
\ENDWHILE
\end{algorithmic}
\label{algo:ppath}
\end{algorithm}

\subsection{Path-finding Algorithms as \textit{IRS}}\label{subsec:pathfinding}
Given the problem setup of \textit{IRS}, it is natural to consider the influence path generation process as a path-finding problem, namely, finding a path between two vertices on a graph that minimizes the cost. 
Since the original dataset is not a graph structure, we build an undirected graph from the user-item interaction sequences following the previous practice \cite{wang2018billion}.
First, we map each item in the training dataset to a graph vertex. 
Then, we assign an edge to two vertices if the corresponding items appear consecutively in an interaction sequence and assign equal weight to each edge. 
For simplicity, we select the last item $i_h$ in the viewing history as the user's recent interest and find a path between $i_h$ and the objective item $i_t$ using a path-finding algorithm such as Dijkstra.
We name this framework as \textit{Pf2Inf}.

\begin{figure}[tb]
	\centering
	\includegraphics[width=0.4\textwidth]{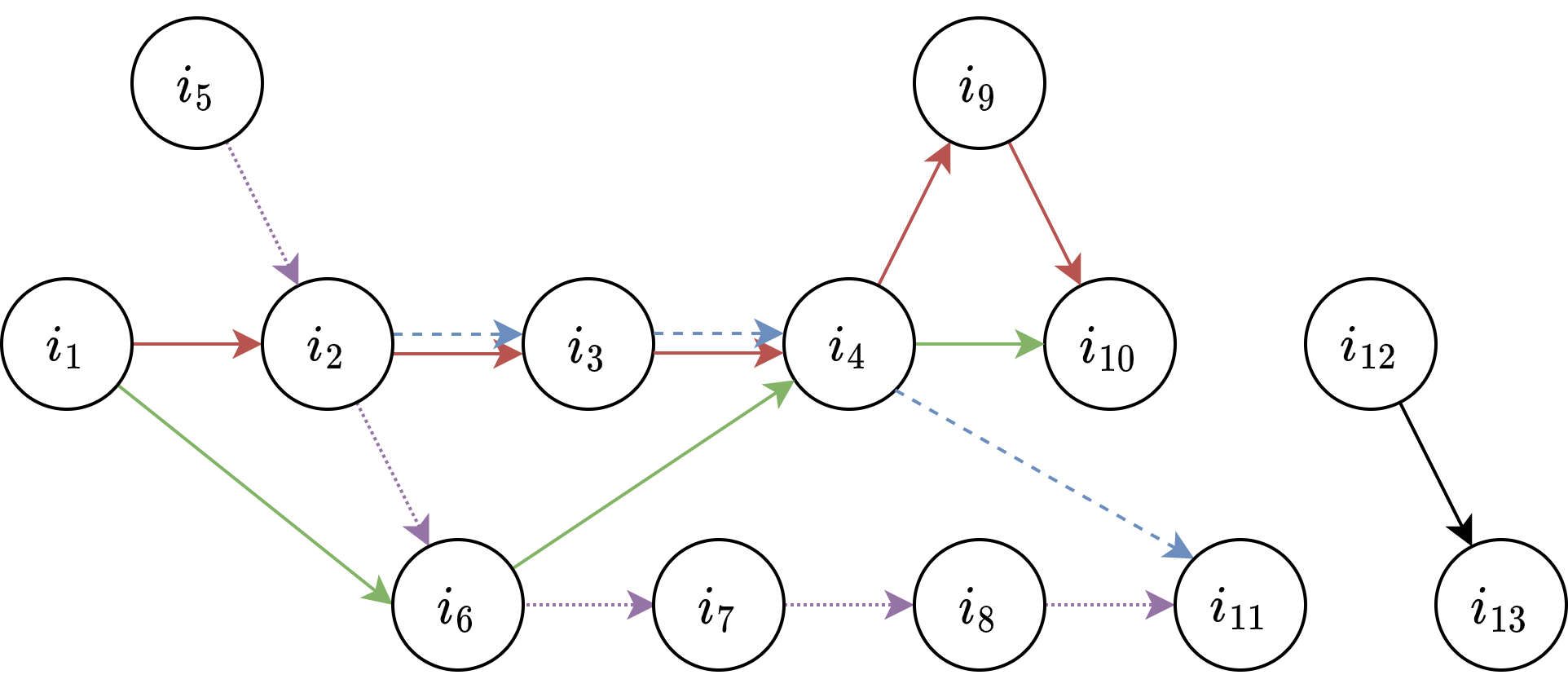}
	\caption{An example of \textit{Pf2Inf}. Here we construct an undirected graph from $5$ user-item interaction sequences (one color per user). The arrow indicates the item order within a sequence. }
	\label{fig:pf_model}
\end{figure}

Figure \ref{fig:pf_model} illustrates an example of \textit{Pf2Inf}. Given a new interaction history $s_h$ ending at $i_1$ and the objective item $i_{11}$, \textit{Pf2Inf} generates an influence path $s_p = \{i_1, i_6, i_4, i_{11}\}$ based on a shortest path algorithm.

\subsection{Adapting Existing \textit{RS} with Greedy Search}
\label{subsec:rec2pros}

Despite its simple architecture, \textit{Pf2Inf} does not take into account the sequential patterns among items in the user's viewing history.
Neither does it work for sparse, disjoint graphs which are likely to exist in the sparse recommendation dataset. For example, in Figure~\ref{fig:pf_model}, given the interaction history $s_h = \{i_3, i_4, i_{10}\}$, it fails to find a reasonable path to the objective item $i_{12}$ by traversing the graph. 



An alternative is to adapt an existing \textit{RS} to \textit{IRS} using a greedy search strategy. Specifically, we can generate an influence path $s_p$ with the following process: given user $u$'s viewing history $s_h$, an existing \textit{RS} can be used to generate the top-$k$ recommendations $\mathcal{R}_k$, which can then be re-sorted based on their distances to the objective item $i_t$ and the closest item is greedily selected into $s_p$. The recommender repeats the process to generate the whole $s_p$. The distance between two items can be calculated based on their embeddings (e.g. obtained from \textit{item2vec} \cite{he2017neural}). 
With this simple strategy, any existing \textit{RS} can be adapted to generate the influence path, and we name this solution as \textit{Rec2Inf}. 

Depending on the backbone \textit{RS}, the \textit{Rec2Inf} framework can enable a recommender system to model user interests and sequential dependencies among items, and approach the objective item at the same time. However, it still suffers from the following limitations: (1) an item is greedily selected into the influence path in each step. The local optimal selections may not ultimately reach the global optimal influence path. For example, if the objective item $i_t$ deviates from the current user interests significantly and all candidate items in $\mathcal{R}_k$ have large distances from $i_t$, $i_t$ will have little effect in selecting the recommended items, hindering the influence path $s_p$ from reaching $i_t$. (2) The objective item $i_t$ is not involved in the training process, which prevents the \textit{RS} from planning early for reaching the goal of $i_t$ in the entire recommendation process.



\subsection{Influential Recommender Network (\textit{IRN})}
\label{subsec:sprs}

\begin{figure}[tb]
	\centering
	\includegraphics[width=0.47\textwidth]{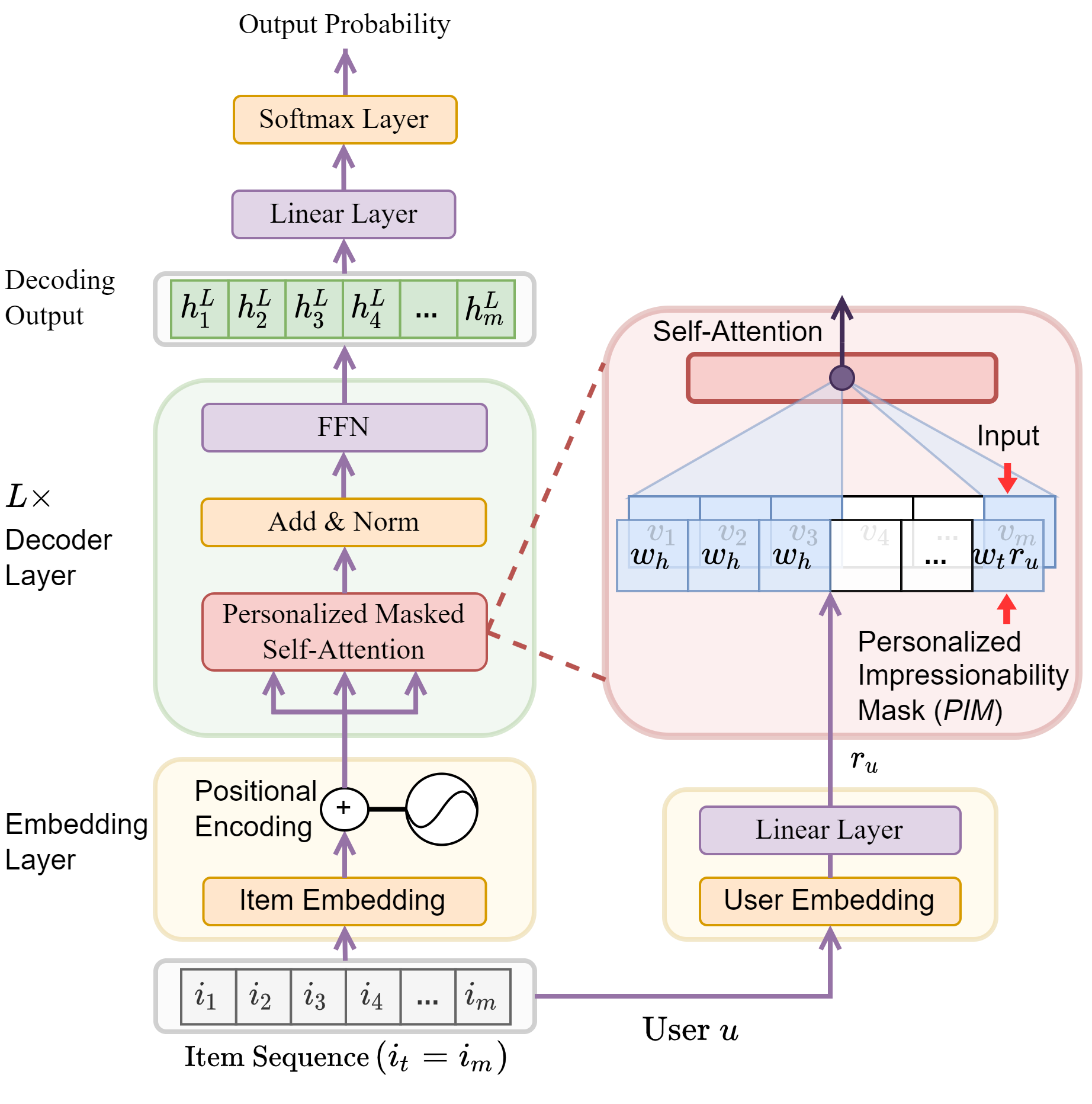}
	\caption{Overall framework of \textit{IRN}. Given a sequence $s = \{i_1, ..., i_m\}$ where the last item is the objective item ($i_t = i_m$), \textit{IRN} first embeds $s$ into vectors. Then it learns the hidden representation of the input sequence with a stack of $L$ decoder layers. The self-attention layers in the decoders incorporate a specially designed \textit{Personalized Impressionability Mask} to cater to users with different acceptance of external influence.}
	\label{fig:overall_prsnn}
\end{figure}

To mitigate the challenges of the \textit{IRS} framework, we propose \textit{Influential Recommender Network} (\textit{IRN}), a neural network model which progressively recommends items to persuade a user to approach the objective. \textit{IRN} can learn the sequential patterns across item sequences, encode the characteristic of the objective item, and consider personalized impressionability at the same time. As shown in Figure~\ref{fig:overall_prsnn}, \textit{IRN} is composed of three components: an embedding layer that vectorizes the input sequence, a stack of $L$ decoder layers, each of which comprises a self-attention layer with personalized masking, and an output layer that projects the hidden states of the decoder into probability distributions over items.

\subsubsection{\textbf{Item Embedding}}
As items are originally represented as discrete tokens, we start by embedding them into dense vectors, which can be processed by deep neural networks. We convert items into both token and position embeddings, which, respectively, represent the context and the position of an item in the sequence. 
Given a sequence of $m$ items~$s = \{i_1, ...i_m\}$, we convert it to a sequence of $d$-dimensional vectors $\mathbf{e}(i_1), ...,\mathbf{e}(i_m)$, with each $\mathbf{e}(i_j)\in R^d$ denoting the embedding for the j-th item, namely,
\begin{equation}
    \mathbf{e}(i_j) = ~TE[i_j] + PE[j]~\mathrm{for}~j = 1, ..., m
\end{equation}
where $TE\in R^{|I|\times d}$ denotes the token embedding matrix and $PE\in R^{m\times d}$ denotes the positional encoding matrix, and "+" refers to addition. 
Inspired by the well-known finding in \textit{NLP} that better initial weights of token embedding can significantly improve the ultimate model performance~\cite{qiu2020pre}, we use pre-trained embedding generated by the \textit{item2vec} model~\cite{barkan2016item2vec}. 

\subsubsection{\textbf{Decoder}}
A Transformer decoder takes the embedding sequence as input and sequentially generates items of the influence path based on the access history and the objective item. 
The decoder maintains a sequence of hidden states $\mathbf{H}= \mathbf{h}_1,...,\mathbf{h}_m$ for the items in the sequence. The hidden state for item $i_j$ is computed based on the items before $i_j$ and the objective $i_t$:
\begin{equation}
    \mathbf{h}_j = \mbox{Dec} (\mathbf{h}_{<j}, \mathbf{h}_t)
\label{equ:dec}
\end{equation}
where $\mathbf{h}_j$ denotes the hidden state of $i_j$. 
The decoder consists of a stack of $L$ self-attention layers with each layer updating the hidden states based on the previous layer. 
\begin{equation}
    \mathbf{H}^{l} = \mbox{Self-Attn}(\mathbf{H}^{l-1})
    \label{equ:dec_wo_u}
\end{equation}
By stacking multiple decoder layers, the model is capable of learning more complicated patterns across the item sequence.
In particular, the first decoder layer takes as input the embedding sequence as the initial hidden states: $\mathbf{H}^0 = \mathbf{e}(i_1), ...,\mathbf{e}(i_m)
$. Here, the self-attention layer performs the layer-wise transformation of the hidden states through the attention mechanism. It takes the hidden states $\mathbf{H}$ as the input for query, key, and value to produce the aggregated value:
\begin{equation}
   \begin{split}
    & \mbox{Self-Attn}(\mathbf{H}) = \mbox{Attention}(\mathbf{H}, \mathbf{H}, \mathbf{H})\\
    & \text{Attention} (\mathbf{Q}, \mathbf{K}, \mathbf{V}) = \text{softmax}(\mathbf{QK}^T/\sqrt{d_k})^T\mathbf{V}
    \end{split}
\end{equation}
where $\mathbf{Q}, \mathbf{K}, \mathbf{V}$ denote the query, key, and value vectors, respectively. Specifically, we use multi-head attention, which is able to attend to information from different representation subspaces at different positions~\cite{li2018multi,sun2019bert4rec, kang2018self}. To enable the decoder to be aware of the objective item, we extend the self-attention layer with a \textit{Personalized Impressionability Mask} (\textit{PIM}). The \textit{PIM} determines (1) which items can be seen by the attention layer, and (2) to what extent the item will be aggregated to produce the hidden states.
We will discuss the two effects of \textit{PIM} in the following sections.


\begin{figure}[tb]
	\centering
	\includegraphics[width=0.47\textwidth]{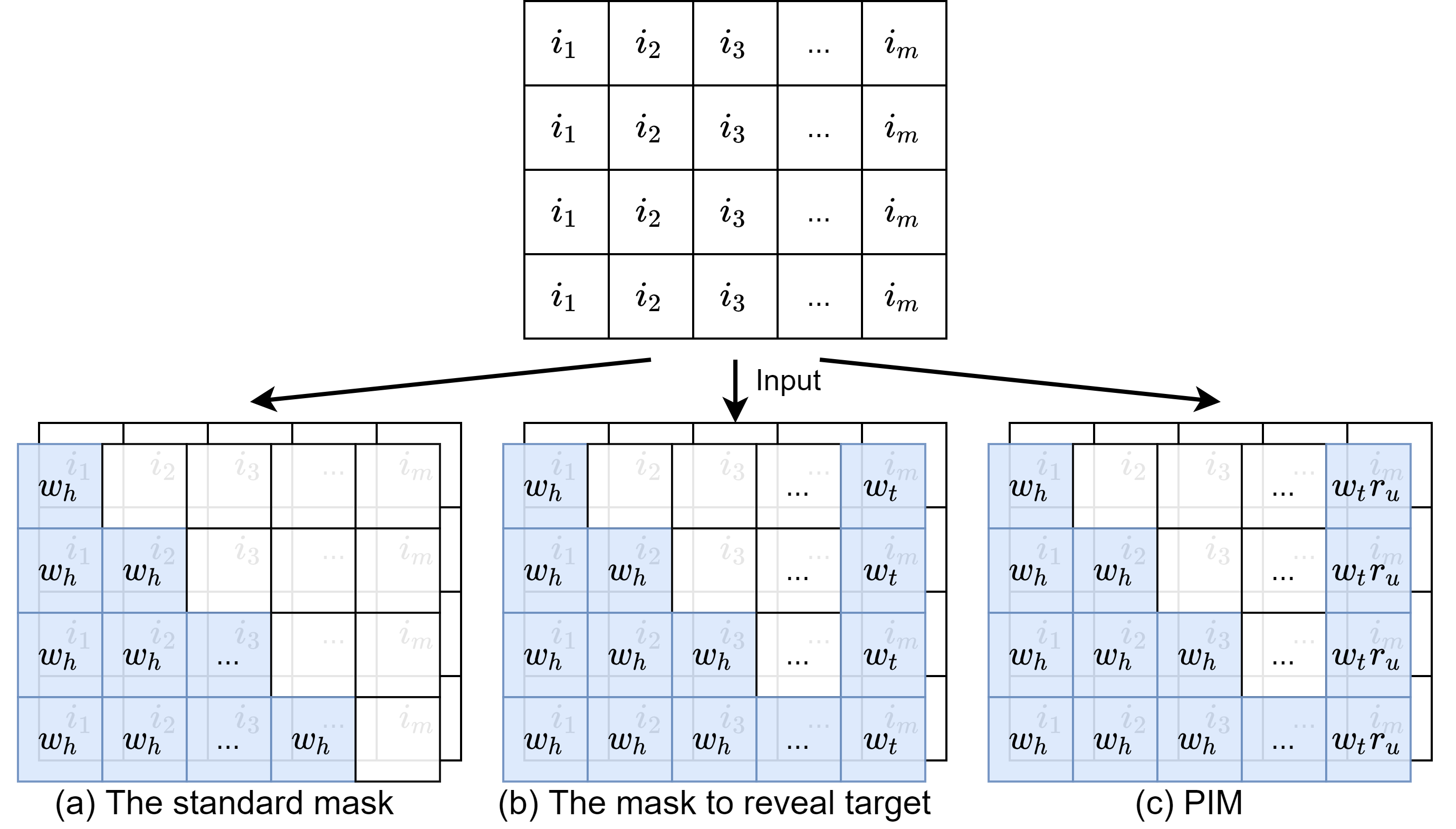}
	\caption{Personalized Impressionability Mask (PIM). $\{i_1,..., i_m\}$ is the input sequence, where the last item $i_m$ is the objective item. (a) displays the standard mask of the decoder. For example, at the second line where the model predicts $i_3$, only $i_1$ and $i_2$ (in blue squares) are perceived by the attention layer while the follow-up items are masked. (b) displays the special mask to reveal the objective, and (c) displays the \textit{PIM} decided by the personalized impressionability factor $r_u$.}
	\label{fig:mask}
\end{figure}

\subsubsection{\textbf{Perceiving objective}}
\label{subsubsec:p_attention_mask}
\textit{PIM} is an extension of the standard attention mask in that it enables the decoder to perceive not only previous items but also the objective item. Figures~\ref{fig:mask}(a) and \ref{fig:mask}(b) illustrate the idea. The standard Transformer decoder utilizes a causal (triangular) attention mask to prevent future items from participating in the prediction. Based on the causal attention, \textit{PIM} introduces attention weights to the objective (Figure~\ref{fig:mask}(b)). This enables the attention layer to be aware of the objective item $i_t$ in addition to the preceding items so that it can take in the extra effect of $i_t$ in generating the hidden states. 

Furthermore, we propose a weighted attention mask to reflect the influence of the objective item $i_t$ ($i_t = i_m$). As shown in Figure~\ref{fig:mask}, each item in the input sequence corresponds to a mask weight (i.e. blue square), which will be added to calculate the attention weight inside the decoder. The item with a larger mask weight may contribute more since its corresponding attention weight is larger. Consequently, the item will be regarded as more relevant in generating the output. Let $w_t$ denote the mask weight for the objective item and $w_h$ denote the mask weight for the preceding history. We enforce the model to pay more attention to the objective item by setting $w_t > w_h$. Additionally, by varying the scale of $w_t$, we can control the aggressiveness degree of \textit{IRN} in performing influential recommendations.

\subsubsection{\textbf{Personalized Impressionability Factor}}
Users have personalized curiousness of exploration \cite{zhao2016much}, which may affect their acceptance of external influence. For an impressionable user, \textit{IRN} can be more aggressive and assign more weight to the objective, while for users who have focused interests, \textit{IRN} should be less aggressive in approaching the objective and consider more the user's historical interests. 
We define the \textit{Impressionability} that measures the users' acceptance of persuasion and utilize the \textit{Personalized Impressionability Factor}, denoted by $r_u$, to control the aggressiveness of influential recommendation. $r_u$ is integrated into the mask weight so that the self-attention model can have varying strategies for different users. $r_u$ is computed as:
\begin{equation}
\begin{aligned}
 &\mathbf{e}(u) = ~\mathcal{U}[u] \\
 & r_u = ~\mathbf{W}^U\mathbf{e}(u)
\end{aligned}
\end{equation}
where $\mathbf{e}(u)$ denotes the embedding of $u$ with dimension $d'$, $~\mathcal{U}$ denotes the user embedding matrix, and $\mathbf{W}^U \in \mathbb{R}^{1 \times d'}$ denotes the parameters of the linear transformation. 
The factor $r_u$ is multiplied by the initial attention mask weight for the objective item $w_t$, resulting in the final attention weights (Figure~\ref{fig:mask}(c)). By integrating $r_u$ into the decoder, Equations~\ref{equ:dec} and \ref{equ:dec_wo_u} are extended as:
\begin{equation}
   \begin{split}
    & \mathbf{h}_j = ~ \mbox{Dec}(\mathbf{h}_{<j}, \mathbf{h}_t, r_u)\\
    & \mathbf{H}^{l} = ~ \mbox{Self-Attn}(\mathbf{H}^{l-1}, r_u),~\mathrm{for}~l=1, ...,L
    \end{split}
\end{equation}

\subsubsection{\textbf{Pre-padding Input Sequences}}
The original item sequences can have varying lengths. Therefore, to accommodate the \textit{PIM}, we apply pre-padding on the input sequence instead of the widely used post-padding scheme~\cite{dwarampudi2019effects}. For example, if the maximum sequence length is 5, then the pre-padded sequence for $\{i_1, i_2, i_3\}$ is $\{PAD, PAD, i_1, i_2, i_3\}$, while the post-padded sequence is $\{i_1, i_2, i_3,\allowbreak PAD, PAD\}$, where $PAD$ is the padding token. Pre-padding allows the objective item to stay at a fixed position (i.e., the last item in the sequence) whereas post-padding has different positions for the objective item depending on the sequence length. The fixed position of the objective item avoids extra noise brought to the model due to the variable-length input sequences in position encoding and accelerates the computation~\cite{li2021easy}.

\subsubsection{\textbf{Output Probability}}
Finally, we project the decoder output $\mathbf{H}^L = \{\mathbf{h}_1^L, ..., \mathbf{h}_m^L\}$ to the item space with a linear layer and generate the output probability using the softmax function. Formally, the probability distribution for generating the $j$-th item can be computed as:
\begin{equation}
    \begin{aligned}
    p(i_j|i_{<j}, i_t, u) = \mbox{softmax}(\mathbf{W}^p \mathbf{h}_{j-1}^L)
    \end{aligned}
\end{equation}
where $W^p \in \mathbb{R}^{|I| \times d}$ is the projection matrix.
\textit{IRN} recommends an influence path by generating items one by one according to the probability distribution $p(i_j|i_{<j}, i_t, u)$ at each step. 

\subsubsection{\textbf{Objective Function}}
Given a sequence of items $s = \{i_1, ..., \allowbreak i_m\}$ for user $u$ ($s \in S_u$), where the last item is selected as the objective item ($i_t = i_m$). Assuming that $u$'s interests start at $i_1$, the actual path from $i_1$ to $i_m$ should have maximized probability, which is equivalent to minimizing the conditional perplexity:
\begin{equation}
\begin{aligned}
    \mathrm{PPL}(i_1,...,i_m|i_t, u) =&~ P(i_1,...,i_m|i_t, u)^{-\frac{1}{m}}\\
    =&~\bigg(\prod_{j=1}^m P(i_j|i_{<j}, i_t, u)\bigg)^{-\frac{1}{m}}
    \label{equ:cond_pp}
\end{aligned}
\end{equation}
Apart from items preceding $i_j$ in the sequence, the conditional probability of $i_j$ considers both the objective item $i_t$ and the personalization of $u$. 
To avoid numerical underflow, we minimize its logarithmic form instead, which is equivalent to minimizing the cross-entropy loss function:
\begin{equation}
    \mathcal{L} = \frac{1}{|S|}\sum_{u\in U}\sum_{s\in S_u} \bigg(-\frac{1}{m}\sum_{j=1}^m \log P(i_j|i_{<j}, i_t, u)\bigg)
\end{equation}
where $S = \mathop{\bigcup}_{u=1}^{|U|} S_u$. It is worth noting that each sequence $s=\{i_1,...,\allowbreak i_m\}$ belongs to one user and one user may have multiple sequences.

\section{Experiment}
\label{sec:experiment}

We evaluate the proposed framework using two commonly used recommendation datasets. 
To meet the requirement of \textit{IRS}, we design several new metrics to assess the model performance.  
Since \textit{IRS} recommends an item sequence, many sequence-item interactions may not exist in the original dataset, making it difficult to evaluate the user's acceptance of these items.
To mitigate the problem, we use a trained \textit{evaluator} to provide probability measures for the unknown sequence-item interactions. 
We will explain in detail the experiment design and outcomes in the remaining sections.  

\begin{table}[tb]
    \centering
    \caption{Statistics of Datasets after Preprocessing. }
    \resizebox{0.45\textwidth}{!}{
    \begin{tabular}{cccccc}
    \toprule
    \multirow{2}{*}{Dataset}&\multirow{2}{*}{Users}&\multirow{2}{*}{Items} &\multirow{2}{*}{Interactions}&\multirow{2}{*}{Density} &Avg. Items \\
    &&&&&per User\\
    \midrule
    Lastfm&896&2,682&28,220&1.17\%&31\\
    Movielens-1m&6,040&3,415&996,183&4.83\%&164\\
    \bottomrule
    \end{tabular}}
    \label{tab:datasets_info}
\end{table}

\subsection{Datasets}
Since online user-item interaction datasets with feedback from real users are not available, we design experiments based on offline datasets that are suitable for the influential recommendation. Specifically, we consider datasets that satisfy the following properties : (1) user-item interactions should exhibit sequential patterns and interaction timestamps are available, and (2) cost in user-item interactions should not be too high to be practical. 
For example, e-commerce purchase datasets may not be suitable because promoting the objective item by asking the user to buy other items is less practical.  
Taking these requirements into consideration, we select two widely used datasets, namely, 
Movielens-1m\footnote{https://grouplens.org/datasets/movielens/1m} and
Lastfm\footnote{https://grouplens.org/datasets/hetrec-2011/} which have timestamps indicating when the user-item interactions take place. 
Moreover, watching movies and listening to music have relatively low costs. 

 
\subsubsection{\textbf{Preprocessing}}
Following the common practice \cite{he2016fusing,wu2016collaborative}, we consider all numeric ratings (tagging behavior) as positive feedback. 
For both datasets, we group the interaction records by the user and order the sequence by interaction timestamps. 
For the Lastfm dataset, we combine multiple consecutive behaviors of the same user-item pair into one single interaction. 
To better reveal the sequential patterns encoded in the dataset, we follow the common practice \cite{rendle2010factorizing,kang2018self, zhang2019feature, sun2019bert4rec} and filter out users and items with less than $5$ interactions.
Table~\ref{tab:datasets_info} shows the statistics of the final dataset.


\subsubsection{\textbf{Dataset Splitting}}\label{subsec:dataset_split}
We split the dataset into training and testing sets through the following steps. 
Given a user $u$'s complete viewing history $s=\{i_1,...,i_{q-1},i_q\}$, we retain the last item $i_q$ for the next-item evaluation task in testing and randomly split the remaining part into multiple continuous, non-overlapping subsequences $\{\{i_{1, 1},...,i_{1, m_1}\},... ,\{i_{k, 1}, \allowbreak ...,i_{k, m_k}\},\allowbreak ...\}$, where each subsequence's length $m_k$ is between $l_{min}$ and $l_{max}$.
For viewing history that is shorter than $l_{min}$, we prolong its length to $l_{min}$ through padding.
We set $l_{min} = 20$, $l_{max}=50$ for Lastfm and $l_{min} = 20$, $l_{max}=60$ for Movielens-1m, respectively. 
The last item $i_{k, m_k}$ of each subsequence is selected as the objective item $i_t$ during the training of \textit{IRN}. 
All subsequences of a user $u$ constitute the user's sequence set $S_u$, and the model training and validation are based on $S = \mathop{\bigcup}_{u=1}^{|U|} S_u$. 
To construct the testing set, we obtain the starting user behavior sequence $s_h=\{i_1, ..., i_{q-1}\}$ for each user. 
Based on the $s_h$ and an objective item $i_t$, each \textit{IRS} framework generates the influence path $s_p =\{i_j, i_{j+1}, ...\}$ following the process illustrated in Figure~\ref{fig:eval_setting}. 
We will further explain the criteria for choosing $i_t$ in Section \ref{subsec:objective_selection}.
Finally, we obtain a testing set of size $|U|$. 
It is worth noting that the test set does not contain ground-truth paths and each \textit{IRS} model generates influence paths based on the same test set independently.


\subsection{\textit{IRS} Evaluation}
\label{subsec:evaluation}
\subsubsection{\textbf{Objective Item Selection}}\label{subsec:objective_selection}
As introduced in previous sections, the objective item $i_t$ can influence the \textit{IRS}'s strategy when generating the influence path $s_p$. For the sake of fairness, for each $s_h$ in the testing set, we randomly sample an objective item $i_t$ from the whole item space based on the following criteria:
(1) $i_t \not\in s_h$. Under the problem formulation of \textit{IRS}, the objective should be a new item to the user.
(2) $i_t$ has at least $5$ interaction records. This prevents unpopular items from being selected as the objective item.

\subsubsection{\textbf{Evaluation Metrics}}
Different from the traditional ``next-item'' \textit{RS} which only needs to evaluate the single recommended item based on a user's viewing history, \textit{IRS} has to evaluate the influence path (i.e., a sequence of recommended items), where both the user acceptance and the path's smoothness need to be taken into account. For example, a long influence path may turn out to be continuously related to the user's current interest but fail to reach the objective item, while a short influence path may recommend the objective item directly but deviates too much from the user's interest and is therefore rejected by the user. 

In order to measure a user's interest over the influence path $s_p$, or whether $s_p$ can better lead the user towards the objective item $i_t$, we need to estimate the joint probability $P(s_p|s_h)$ as well as the conditional probability $P(i_t|s_p)$. This can be formulated as a language model which defines the probability of each sequence appearing in a sequence corpora:
\begin{equation}
    \begin{aligned}
        P(i_1...i_q) =~& P(i_1)P(i_2|i_1)...P(i_q|i_{<q}) = \prod_{k=1}^qP(i_k|i_{<k})
    \end{aligned}
\end{equation}

For these concerns, we evaluate the influence path $s_p$ from two aspects: 
(1) whether the influence path $s_p$ can successfully shift the user interests towards the objective $i_t$,
and (2) whether $s_p$ satisfies the user's interest.
We design and employ the following metrics:
\begin{itemize}[leftmargin=0.4cm]
    \item \textbf{Success Rate ($SR_{M}$)} measures the ratio of generated $s_p$'s that can successfully reach the objective item $i_t$ within the maximum length $M$ (as denoted in Algorithm \ref{algo:ppath}):  
    \begin{equation}
        SR_{M} = \frac{1}{|U|}\sum_{u=1}^{|U|} 1(i_t^u \in s_p^u) 
    \end{equation}
    \item \textbf{Increase of Interest ($IoI_{M}$)} measures the change of user's interest in the objective item $i_t$ after being persuaded through $s_p$:
    \begin{equation}
        IoI_{M} = \frac{1}{|U|}\sum\limits_{u=1}^{|U|} \bigg(\log P(i_t^u|s_h^u \oplus s_p^u) - \log P(i_t^u|s_h^u) \bigg)
    \end{equation}
    where ``$\oplus$'' refers to the concatenation of two sequences; $P(i_t|s)$ is the probability that the objected item $i_t$ is accepted by a user with viewing sequence $s$. Here we use log form probability to avoid numerical underflow.
    \item \textbf{Increment of Rank ($IoR_{M}$)} measures the increase of the ranking for the objective item $i_t$ after being persuaded through $s_p$:
    \begin{equation}
        IoR_{M} = \frac{1}{|U|}\sum\limits_{u=1}^{|U|} -\bigg(R(i_t^u | s_h^u \oplus s_p^u) - R(i_t^u | s_h^u) \bigg)
    \end{equation}
    where $R(i_t^u | s_h^u)$ denotes the ranking of the objective item $i_t^u$ based on $P(i_t^u | s_h^u)$. 
    \item \textbf{Perplexity (PPL)} measures the naturalness and smoothness of the influence path $s_p$, that is, how likely the path can appear in the viewing history. In our experiment, PPL is defined as the conditional probability of that $s_p$ follows $s_h$: 
    \begin{equation}
        \begin{aligned}
            \mathrm{PPL}(s_p|s_h) = & \bigg( \prod_{k=1}^{|s_p|}P(i_k|s_h \oplus i_{<k}) \bigg)^{-\frac{1}{|s_p|}}\\
            \log(\mathrm{PPL}) = &\frac{1}{|U|}\sum_{u=1}^{|U|} \log \mathrm{PPL}(s_p^u|s_h^u)     \\
             = & \frac{1}{|U|}\sum_{u=1}^{|U|} \sum_{k=1}^{|s_p^u|}\log P(i_k^u|s_h^u \oplus i_{<k}^u)
        \end{aligned}
    \label{equ:n_gram}
    \end{equation}
    where $s_h \oplus i_{<k}$ denotes the concatenated sequence of the viewing history $s_h$ and the preceding items of $i_k$ in $s_p$, and $|s_p|$ denotes the length of $s_p$. Note that the logarithm form of perplexity is equivalent to the average of the log probability that the user accepts the path item $i_k$ in each step of the influence path. 
\end{itemize}


\subsubsection{\textbf{IRS Evaluator}}
As introduced above, the computation of $IoI_M$, $IoR_M$, and PPL relies on the term $P(i|s)$, which refers to the probability that item $i$ is accepted by a user given the viewing history $s$. This presents a unique challenge for the offline evaluation of \textit{IRS} with the logged dataset: many sequence-item interactions may not exist in the users' viewing histories, especially when the sequence $s$ is long. 

Traditional \textit{NLP} approaches estimate the term $P(i|s)$ by counting the occurrences of items as follows:
\begin{equation}
    P(i|s) = \frac{C(s \oplus i)}{C(s)}
\end{equation}
where $C(s)$ refers to the times that sequence $s$ occurs in the sequence corpora and $s \oplus i$ refers to the new sequence with $i$ appended into $s$. 
Due to the sparsity of \textit{RS} datasets, there can be many zero counts (i.e., $P(i|s)=0$) in computing the conditional probabilities, which makes the counting approximation approach ineffective. 

Inspired by the common practice of \textit{simulation based} experiments in reinforcement learning based \textit{RS} \cite{chen2019top, Huang_simulator2020, li2015counterfactual, huang2020keeping} and combinatorial recommendation \cite{yue2011linear, wang2019sequential}, we adopt an independent \textit{IRS} evaluator $\mathbb{E}$ to tackle the problem.
The evaluator $\mathbb{E}$ measures the relevance of item $i$ by estimating the probability of item $i$ following the sequence $s$, namely,
\begin{equation}
\label{eqn:evaluator}
    P(i|s) = \mathbb{E}(i, s)
\end{equation}
Specifically, we train an independent next-item recommender model which outputs a measurement for the probability term $P(i|s)$.
The recommender model $g$ contains a softmax layer which generates a probability distribution $\mathbf{D}$ over all items given an input sequence $s$. Finally, we compute $P(i|s)$ as:
\begin{equation}
\begin{aligned}
    \mathbf{D}(s) &= \mathbb{E}(s) = \mbox{Softmax}(g(s)) \\
    P(i|s) &= \mathbb{E}(i, s) = \mathbf{D}(s)[i]\\
\end{aligned}
\end{equation}
Although the evaluator is an approximation of real users, it provides a reasonably fair offline evaluation strategy for sequence-item interactions that do not appear in the dataset. 

An accurate evaluator should be capable of capturing the sequential patterns encoded in the viewing sequence and recommend the next item accurately. Therefore, we experiment with multiple state-of-the-art sequential recommenders and select the optimal one as the final evaluator. We measure the performance of the evaluator using \textbf{HR@20} (Hit Ratio) and \textbf{MRR} (Mean Reciprocal Rank):
\begin{equation}
\begin{aligned}
    HR@20 & =  \frac{1}{|U|}\sum_{u=1}^{|U|} 1\big(R(i_q^u | s_h^u) \leq 20 \big)\\
    MRR & = \frac{1}{|U|}\sum_{u=1}^{|U|} \frac{1}{R(i_q^u | s_h^u)}\\
\end{aligned}
\label{eqn:hr20}
\end{equation}
where $|U|$ denotes the size of the testing set, $s_h^u$ the viewing history sequence (not containing $i_q^u$) of user $u$, and $R(i_q^u | s_h^u)$ the ranking of item $i_q^u$ in each \textit{IRS}; $1(*)$ equals to $1$ if the condition ``$*$'' is true and $0$ otherwise. 
Recall that the last item ($i_q$) of the input sequence $\{i_1,...,i_{q-1},i_q\}$ is the labelled item. 

Specifically, we consider the following models for the \textit{IRS} evaluator:
\begin{itemize}[leftmargin=0.4cm]
    \item \textbf{GRU4Rec}~\cite{hidasi2018recurrent} employs the recurrent neural network (\textit{RNN}) based architecture to model user behavior sequences for sequential recommendation.  
    \item \textbf{Caser}~\cite{tang2018personalized} utilizes the convolutional neural network (\textit{CNN}) to learn the sequential patterns of user-item interactions in both horizontal and vertical way.
    \item \textbf{SASRec}~\cite{kang2018self} models the sequential dynamics of recommendation by using a self-attention-based model, which allows for capturing both long-term and short-term semantics. 
    \item \textbf{Bert4Rec}~\cite{sun2019bert4rec} further improves the user behavior sequences modeling with bidirectional self-attention and is considered one of the strongest baselines for the sequential recommendation.
\end{itemize}
The original models were implemented in various deep learning frameworks, and we port them to PyTorch based on the originally published code\footnote{https://github.com/graytowne/caser\_pytorch}\footnote{https://github.com/hidasib/GRU4Rec}\footnote{https://github.com/kang205/SASRec}\footnote{https://github.com/FeiSun/BERT4Rec}. We tune all hyperparameters of these models using the validation set and compare their results on the testing set.

\begin{table}[tb]
\renewcommand{\arraystretch}{1.3}

    \centering
    \caption{Performance of IRS Evaluator. The bold number indicates the best performance and the underlined number indicates the second best performance. }
	\begin{tabular}{l|ll|ll}
		\hline
		\multirow{ 2}{*}{\diagbox{Method}{Dataset}}&\multicolumn{2}{c|}{Lastfm}&\multicolumn{2}{c}{Movielens-1M}\\ \cline{2-5}
	    Method&$HR@20$&$MRR$&$HR@20$&$MRR$\\ \hline\hline
	    {GRU4Rec}&0.0392&0.0081&0.243&\underline{0.063}\\ \hline
	    {Caser}&\underline{0.0460}&\underline{0.0106}&0.252&0.062\\ \hline
	    {SASRec}&0.0445&0.0097&\underline{0.257}&0.062\\ \hline
	    {Bert4Rec}&\textbf{0.0489}&\textbf{0.0199}&\textbf{0.264}&\textbf{0.076}\\ \hline
	\end{tabular}
	\label{tab:evaluator}
\end{table}

Table \ref{tab:evaluator} shows the comparison results of \textit{IRS} evaluators. Among the tested sequential recommender models, \textit{Bert4Rec} has achieved the best performance in terms of all evaluation metrics, with around $4.5\%$ $HR@20$ and $58.4\%$ $MRR$ improvements against the second-best model. Therefore, \textit{Bert4Rec} is selected as the evaluator and used to provide a probability measure of $P(i|s)$ in the following experiment discussion.

\begin{table*}[tb]
\renewcommand{\arraystretch}{1.3}
	\caption{Comparison of various models on two datasets ($M = 20$). ``Vanilla'' means the original version of baselines, while ``Pf2Inf'' and ``Rec2Inf'' denote baselines that have been adapted to the IRS framework. The bold number indicates the best performance, and the underlined number indicates the second-best performance within IRS approaches. }
	\centering
	\begin{tabular}{l|l||llll||llll}
		\hline
		\multicolumn{2}{c||}{\multirow{ 2}{*}{\diagbox{Method}{Dataset}}}&\multicolumn{4}{c||}{Lastfm}&\multicolumn{4}{c}{Movielens-1M}\\ \cline{3-10}
		\multicolumn{2}{c||}{}&$SR_{20}$&$IoI_{20}$&$IoR_{20}$&$\log$(PPL)&$SR_{20}$&$IoI_{20}$&$IoR_{20}$&$\log$(PPL)\\ \hline\hline
        \multirow{2}{*}{Pf2Inf}&Dijkstra& 0.031& 0.0192& 3.73 & 9.88 & 0.069 & 0.009 &362.8& 6.97\\ \cline{2-10}
        &MST& 0.0016& -0.0531& -11.5 & 8.13 & 0.031 & -0.026 &52.0& 8.47\\ \hline\hline
		\multirow{7}{*}{Vanilla}&POP& 0.00& -0.0119& 9.74 & 7.27 & 0.005 & -0.078 & 18.1 & 5.39\\ \cline{2-10}
		&BPR& 0.001& -0.123 & 5.57& 6.64 & 0.006 & -0.171 & 41.5 & 5.62\\ \cline{2-10}
		&TransRec& 0.00& -0.0720& -0.483& 6.69 & 0.005 & 0.083 & 19.9 & 5.11\\ \cline{2-10}
		&GRU4Rec& 0.001& -0.0818& 4.72 & 6.27 & 0.00 & -0.583 & -7.32 & 4.41 \\ \cline{2-10}
		&Caser& 0.002& 0.0110& 13.6 & 6.71& 0.00 & -0.887 & 7.36 & 3.78\\ \cline{2-10}
		&SASRec& 0.00& 0.0257& -15.0 & 6.36 & 0.001 & -0.003 & -0.918 & 3.95\\ \hline\hline
		\multirow{6}{*}{Rec2Inf}&POP& 0.0234& \underline{0.203} & 47.7 & 7.46  & 0.019 &  1.35 & 238.9 & 5.89 \\ \cline{2-10}
		&BPR& 0.0279 & -0.0218 & 41.9 & 6.80 & 0.021 & 0.930 & 242.8 & 6.01 \\ \cline{2-10}
		&TransRec& \underline{0.0625} & 0.0218 & \underline{48.9} & 6.77 & 0.042 & \underline{1.59} & 508.0 & 5.50\\ \cline{2-10}
		&GRU4Rec& 0.0140 & 0.0392 & 20.8 & \textbf{6.37} & 0.012 & 0.91 & 130.1 & 5.22 \\ \cline{2-10}
		&Caser& 0.0246 & 0.0236 & 15.82 & 6.81 & \underline{0.073} & 1.38 &  \underline{614.0} & \textbf{4.74} \\ \cline{2-10}
		&SASRec& 0.0137 & 0.0205 & 42.7 & 7.78& 0.014 & 1.10 & 139.8 & 6.03 \\ \hline
		\multicolumn{2}{c||}{IRN}& \textbf{0.142} & \textbf{0.572} & \textbf{160.7} & \underline{6.65} & \textbf{0.259} & \textbf{2.61} & \textbf{804.7} & \underline{4.94} \\ \hline
		
    	\end{tabular}
	\label{tab:random_result}
\end{table*}

\subsection{Baselines}
We compare our approach with two path-finding algorithms under the \textbf{\textit{Pf2Inf}} framework and six \textit{RS} baselines under the \textbf{\textit{Rec2Inf}} framework. 
Overall, we use the following approaches as baselines:
\begin{itemize}[leftmargin=0.4cm]
    \item \textbf{Dijkstra} finds the shortest path between two nodes in the item graph (described in Section \ref{subsec:pathfinding}). We select the influence path as the first $M$ items along the path from the last item in $s_h$ to $i_t$.
    \item \textbf{MST} constructs a minimum spanning tree from the constructed item graph. We select the influence path using the same process as \textit{Dijkstra}. 
    \item \textbf{POP} sorts all the items by occurrence and recommends the most popular items to the user. 
    \item \textbf{BPR}~\cite{rendle2012bpr} is a commonly used matrix factorization method that optimizes a pairwise objective function.
    \item \textbf{TransRec}~\cite{he2017translation} models item-item and user-item translation relationships for large-scale sequential prediction.
    \item \textbf{GRU4Rec}~\cite{hidasi2018recurrent} is an \textit{RNN} based model for sequential recommendation.  
    \item \textbf{Caser}~\cite{tang2018personalized} is a \textit{CNN} based model for sequential recommendation. 
    \item \textbf{SASRec}~\cite{kang2018self} is a sequential recommender based on self-attention.
\end{itemize}

For \textit{RS} approaches under the \textit{Rec2Inf} framework, we use the recommended hyperparameters in the original implementation and set the size of the candidate set $k=50$. We calculate the item distance using genre feature vector on Movielens-1m, and \textit{item2vec} \cite{barkan2016item2vec} embeddings on Lastfm, respectively.

\subsection{Experiment Result}

\subsubsection{\textbf{Overall Comparison}}
\label{subsubsec:overall_comp}
Table \ref{tab:random_result} shows the overall results of various models on the two datasets. As we can see, \textit{IRN} significantly outperforms other baselines in terms of $IoI_{20}$, $IoR_{20}$, and $SR_{20}$ on both datasets. 
In the Lastfm dataset, $14.2\%$ of the influence paths generated by \textit{IRN} for the testing set can reach the objective item within $20$ steps.
In the Movielens-1M dataset, $25.9\%$ of testing sequences can successfully lead to the objective item, while the strong baseline method \textit{Caser} can only achieve 7.3\%. 
In terms of PPL, \textit{IRN} achieves the second best results, only next to \textit{TransRec} in Lastfm and \textit{Caser} in Movielens-1m, respectively. 
It indicates that \textit{IRN} does not hamper the smoothness of the influence path while achieving superior influencing performance. 
Note that a trade-off exists between the influencing power of the recommender and the smoothness of the influence path. We will further discuss the aggressiveness of \textit{IRS} in Section \ref{subsubsec:aggressive_pers}.

In comparison, \textit{Dijkstra} under the \textit{Pf2Inf} framework achieves greater influencing performance at the cost of high PPL. This indicates that although path-finding algorithms manage to find a path from the viewing history to the objective, it often fails to satisfy the user's current interests and is likely to be rejected by users in the middle of persuasion. 
We also compare \textit{IRN} with the vanilla versions (i.e., the original model without being adapted to the \textit{Rec2Inf} paradigm) of baselines. 
The influence path $s_p$ of each vanilla model is generated by repeatedly recommending the item with the maximum $P(i|s)$.
Since the vanilla model is driven solely by users' current interest, we can observe a $SR_{20}$ and $IoI_{20}$ approximately equal to zero for most vanilla baselines.
Compared to their vanilla versions, these adapted baselines can achieve a higher $SR_{20}$ and have a  significantly larger $IoI_{20}$ and $IoR_{20}$, which indicates that the \textit{Rec2Inf} frameworks are effective in increasing the user's interest in the objective item. 
However, the adapted baselines generally have a higher PPL. 
It is reasonable because they need to pay extra attention to the objective item at each step of the influence path, frequently resulting in a recommendation that suits the user's interest less. 

In summary, \textit{IRN} outperforms the baselines in most evaluation metrics, which validates its capability of generating natural and smooth influence paths.

\subsubsection{\textbf{Short-term Interests Match}}
In addition to evaluating the influencing power of the recommender models, we also investigate whether they can achieve comparable performance in the task of traditional next-item recommendation. 

Table \ref{tab:short-term} displays the performance of various models in terms of next-item recommendation. 
In general, baseline models under the \textit{IRS} framework achieve suboptimal performance compared to the state-of-the-art next-item recommenders. 
We also observe a $2 \sim 20\%$ performance loss compared to their vanilla versions under the \textit{Rec2Inf} framework. 
The reason could be that the \textit{IRS} needs to shift towards the objective $i_t$ early if $i_t$ is far from the user's current interest. This cause the model to recommend an unfamiliar item to the user, thus reducing the recommendation accuracy. 
Yet \textit{IRN} achieves comparable results with respect to $Hit@20$ and $MRR$ on the two datasets, with around $9.0\%$ $HR@20$ and $9.2\%$ $MRR$ performance loss compared to the best-performed $Bert4Rec$. This indicates that even with the objective item $i_t$, \textit{IRN} can still capture the user's current interests without abruptly heading for $i_t$. 

\begin{table}[tb]
\renewcommand{\arraystretch}{1.3}

    \caption{Performance of IRS in terms of next-item recommendation.}
    \renewcommand{\arraystretch}{1.3}

	\begin{tabular}{l|ll|ll|ll}
		\hline
		\multicolumn{3}{c|}{\multirow{ 2}{*}{\diagbox{Method}{Dataset}}}&\multicolumn{2}{c|}{Lastfm}&\multicolumn{2}{c}{Movielens-1M}\\ \cline{4-7}
	    \multicolumn{3}{c|}{}&$HR@20$&$MRR$&$HR@20$&$MRR$\\ \hline\hline
	    
	    &&{GRU4Rec}&0.0392&0.0081&0.243&0.063\\ \cline{2-7}
	    Next-item&&{Caser}&0.0460&0.0106&0.252&0.062\\ \cline{2-7}
	    RS&&{SASRec}&0.0445&0.0097&0.257&0.062\\ \cline{2-7}
	    &&{Bert4Rec}&0.0489&0.0199&0.264&0.081\\ \hline\hline
	    
	    \multirow{4}{*}{IRS}&&{GRU4Rec}&0.0387&0.0079&0.198&0.047\\ \cline{2-7}
	    &&{Caser}&0.0430&0.0102&0.250&0.061\\ \cline{2-7}
	    &&{SASRec}&0.0437&0.0092&0.207&0.041\\ \cline{2-7}
	    &\multicolumn{2}{c|}{IRN}&0.0458&0.0190&0.242&0.073\\ \hline\hline
	\end{tabular}
	\label{tab:short-term}
\end{table}



	    
	    

\begin{figure}[tb]
	\centering
	\includegraphics[width=0.47\textwidth]{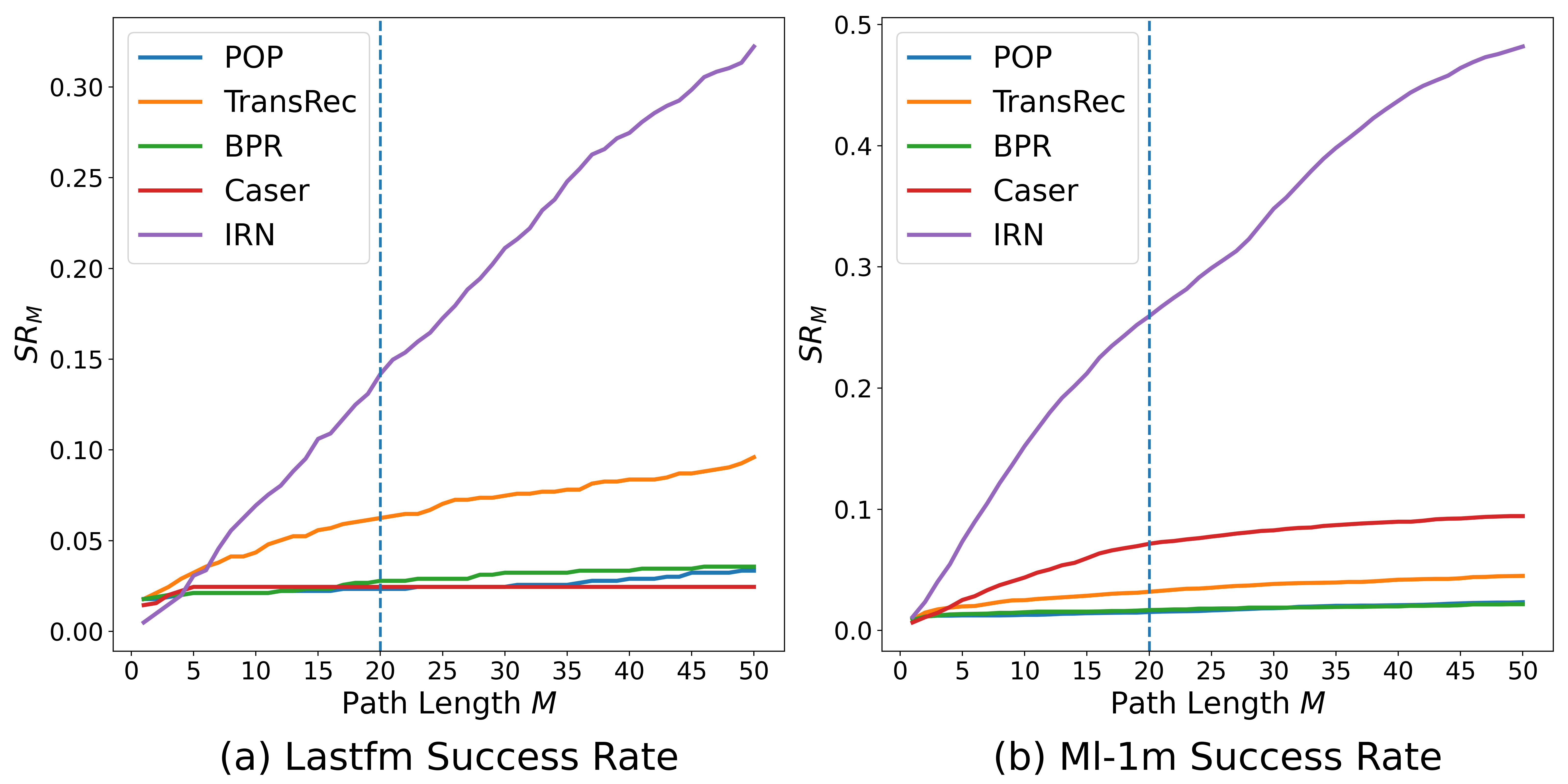}
	\caption{Success rate $SR_M$ versus path length $M$.}
	\label{fig:step}
\end{figure}

\subsubsection{\textbf{Aggressiveness of IRS}}
\label{subsubsec:aggressive_pers}
One notable phenomenon observed in Section~\ref{subsubsec:overall_comp} is that
even though \textit{IRN} significantly outperforms the baselines, the majority of the generated influence paths cannot successfully end at the objective item (e.g., $SR_{20}$ is only 14.2\% and 25.9\% on the two datasets, respectively). 
In this section, we study this phenomenon by analyzing the aggressiveness of influence paths. We hypothesize that whether $s_p$ can reach the objective item is mainly influenced by two factors: the maximum path length $M$ and the \textit{Aggressiveness Degree} (\textit{AD}) of each \textit{IRS}.
Hence, we investigate the relation between these two factors and the success rate of reaching the objective item ($SR$).

Figure~\ref{fig:step} illustrates the relationship between the maximum path length and the success rate. For clarity, we only compare \textit{IRN} to the strong baselines in the \textit{Rec2Inf} framework. 
In general, for all methods, the success rate increases as $M$ increases since longer influence paths increase the chance of reaching the objective. In addition to the outstanding performance, \textit{IRN} achieves a steady improvement of $SR$ as the influence path length increases, whereas the baselines' $SR$ quickly flattens out to a low level. This can be attributed to the fact that \textit{IRN} incorporates the objective item in the training process, thus enabling long-range planning in the influencing process. In contrast, the baseline models make an influential recommendation at each step independently and locally.

\begin{figure}[tb]
	\centering
	\includegraphics[width=0.47\textwidth]{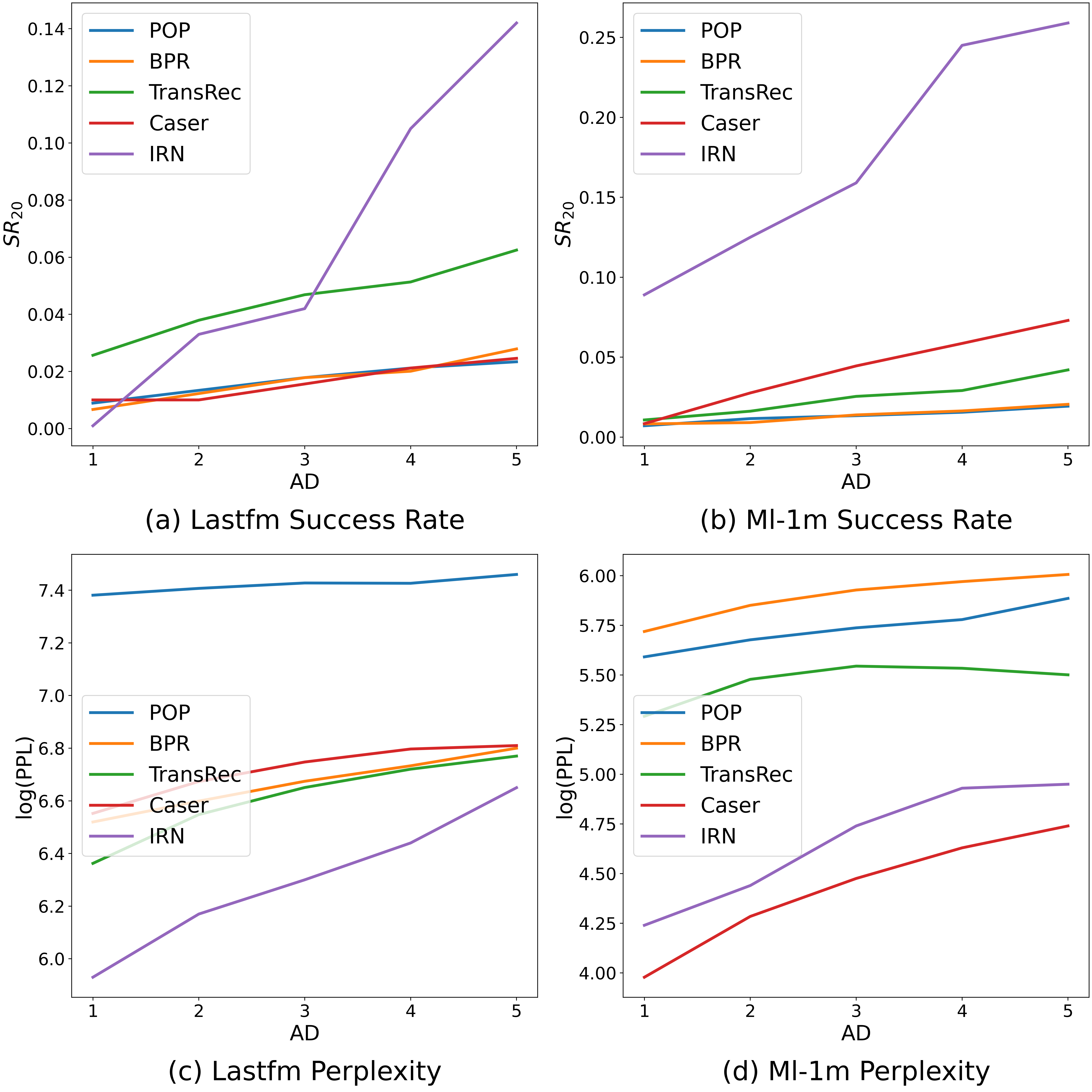}
	\caption{Performance of \textit{IRS} under different aggressiveness degrees.}
	\label{fig:aggressive}
\end{figure}

Figure~\ref{fig:aggressive} illustrates the impact of aggressiveness degree (\textit{AD}) when \textit{IRS} is approaching the objective item. Each recommender has its own parameters to control the \textit{AD}. As we can see, \textit{AD} for the baselines is correlated to the size of the candidate set $k$. As introduced in Section \ref{subsec:rec2pros}, the \textit{Rec2Inf} framework selects the item that is closest to the objective from the top $k$ candidate set. When $k=1$, this is equivalent to the original recommender system which simply predicts the next item without considering the objective. When $k$ is set to the total number of items, it may recommend the objective item directly which has zero distance to itself. 
Different from the baselines, \textit{AD} for \textit{IRN} is related to the objective mask weight $w_t$. Larger $w_t$ enforces the model to rely more on the objective item in the recommendation. 

In order to compare \textit{AD}'s influence to baselines and \textit{IRN}, we need to unify \textit{AD}'s of different methods to the same scale. Therefore, we set 5 levels of $k = \{10,20,30,40,50\}$ for \textit{Rec2Inf} baselines and 5 levels of $w_t = \{0, 0.25, \allowbreak 0.5, 0.75, 1\}$ for \textit{IRN}, both 5 levels are then mapped as the x-axis of Figure \ref{fig:aggressive}. 
Figures~\ref{fig:aggressive}(a) and (b) show the value of $SR_{20}$ under different levels of \textit{AD}s. As expected, when \textit{AD} increases, $SR_{20}$ increases accordingly since more attention on the objective item can increase the chance of reaching the objective. In addition, \textit{IRN} significantly outperforms the baselines given the same \textit{AD}. This indicates that the model is effective in leading users towards the objective. 
Figures~\ref{fig:aggressive}(c) and (d) depict the perplexity over different levels of \textit{AD}s. A higher perplexity indicates a lower likelihood that users accept the influence paths. As we can see, most baseline methods exhibit a clear trade-off between $SR_{20}$ and PPL (higher $SR$ tends to have low PPL), probably because the aggressive influencing strategy increases the abruptness of sequential recommendation. 
In contrast, \textit{IRN} is superior to other models with a significantly higher $SR_{20}$ and a smaller PPL, demonstrating its capability of effectively leading users towards the objective and maintaining a high level of smoothness.
Furthermore, we observe that there does not exist a universal optimal \textit{AD} which can achieve the best $SR_{20}$ and PPL simultaneously. This indicates that the selection of \textit{AD} in the \textit{IRS} framework should be adapted to the requirement of the applications.

\subsubsection{\textbf{Effectiveness of the Personalized Mask Weight}}
\label{subsubsec:personalize_mask_weight}

\begin{figure}[tb]
	\centering
	\includegraphics[width=0.47\textwidth]{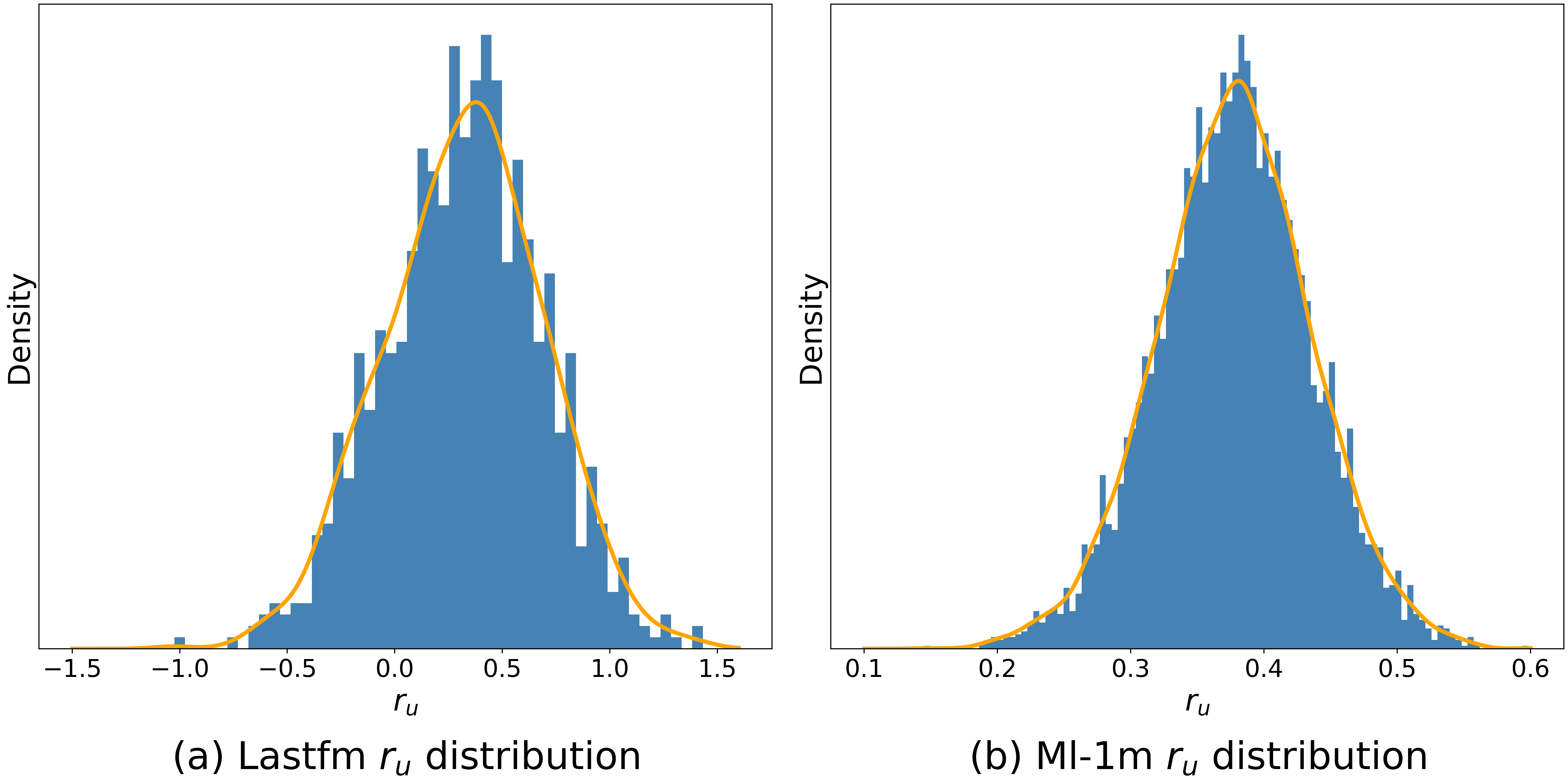}
	\caption{Distribution of the \textit{Personalized Impressionability Factor}.}
	\label{fig:distribution}
\end{figure}

\begin{table}[tb]
\renewcommand{\arraystretch}{1.3}

	\caption{Comparison of Different Masks.}
    \centering
	\begin{tabular}{l|l@{}ll|l@{}ll}
		\hline
		\multirow{ 2}{*}{Mask Type}&\multicolumn{3}{c|}{Lastfm}&\multicolumn{3}{c}{Movielens-1M}\\ \cline{2-7}
	    &$\log(PPL)$\;&$SR_{20}$&$IoI_{20}$&$\log(PPL)$\;&$SR_{20}$&$IoI_{20}$\\ \hline\hline
	    {Type 1}&5.93&0.002&0.011&4.27&0.009&0.054\\ \hline
	    {Type 2}&6.55&0.118&0.477&4.92&0.232&2.01\\ \hline
	    {Type 3}&6.65&0.142&0.572&4.94&0.259&2.61\\ \hline
	\end{tabular}
	\label{tab:compare_mask}
\end{table}

As introduced in Section~\ref{subsec:sprs}, The \textit{Personalized Impressionability Mask} (\textit{PIM}) models a user's personal impressionability about the persuasion. Table \ref{tab:compare_mask} compares the performance of various masking schemes. In ``Type 1'' the model will not pay extra attention to the objective (i.e., $w_h=w_t=0$), in ``Type 2'' the objective has a uniform large weight ($w_t=1$), and in ``Type 3'' the model integrates the \textit{Personalized Impressionability Factor} $r_u$ and the objective has weight $r_uw_t$. Compared to ``Type 1'', ``Type 2'' has significantly improved the user's acceptance of the objective item (measured by $IoI_{20}$ and $SR_{20}$) at the expense of the smoothness of the path (measured by $\log(PPL)$). The addition of $r_u$ further increases the influence performance by approximately $20\%$ without evident impact to the smoothness. 

The \textit{IRN} learns $r_u$ for each user. We further investigate the distribution over $r_u$ values, as shown in Figure \ref{fig:distribution}. A larger $r_u$ indicates that the user is more adventurous and prefers a larger \textit{AD}. Thus, recommended items by \textit{IRN} are more close to the objective. On the other hand, a smaller $r_u$ results in recommendations that depend more on the user's historical interests. In general, $r_u$ follows a normal distribution on both datasets. This validates our intuition that users are indeed different in accepting influential recommendations and that \textit{PIM} is effective in modeling their different personalities.

\subsubsection{\textbf{Stepwise evolution of User Interests}} 
To get a deeper insight into the influencing process, we study the evolution of the influence path in each step. This gives us a micro view of how \textit{IRS} leads the user towards the objective. 
Specifically, we analyze two probabilities, namely, the ``objective probability'' $P(i_t|s_h\oplus i_{<k})$ (i.e., the probability that the user likes objective item $i_t$ at the $k^{th}$ influencing step) and ``item probability'' $P(i_k|s_h\oplus i_{<k})$ (i.e., the probability that the user is satisfied with item $i_k$ at the $k^{th}$ influencing step) over all the influence paths, which can be calculated using the \textit{IRS} evaluator $\mathbb{E}$. 
To facilitate analysis, we average the two probabilities on the testing sequences.  
Note that influence paths may have various lengths due to early reach to the objective, which affects the averaging of $P(i_t|s_h\oplus i_{<k})$ and $P(i_k|s_h\oplus i_{<k})$. Therefore, we exclude early-success paths which account for only a small fraction given $M=20$. 

\begin{figure}[tb]
	\centering
	\includegraphics[width=0.47\textwidth]{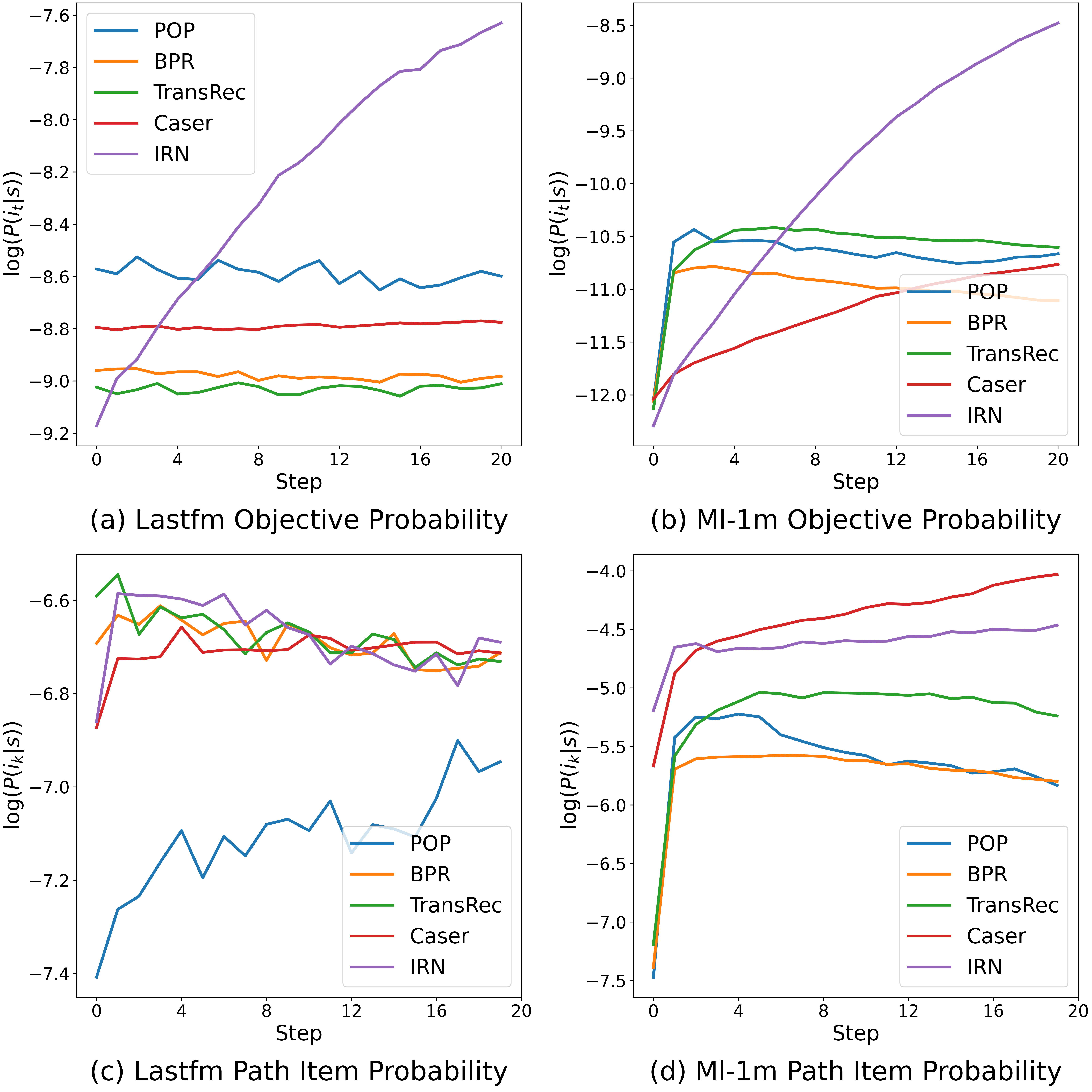}
	\caption{Stepwise evolution of user interests. (a) and (b) display the change of the objective probability as the influence path extends. (c) and (d) display the change of item probability through the path.}
	\label{fig:grad}
\end{figure}

Figure \ref{fig:grad} presents the results of stepwise user interests evolution.
On both datasets, the influence paths generated by \textit{IRN} can lead the user towards the objective steadily while maintaining a high probability to be accepted by the user. 
In comparison, influence paths recommended by most adapted baselines cannot steadily lead users to the objective, as evident from their flat curves. 
In the Movielens-1m dataset, the suboptimal approach \textit{Caser} also demonstrates a steadily increasing probability curve of the user accepting the objective item, but the increase amplitude is far below \textit{IRN}. 
Experiment results show that \textit{IRN} can consistently utilize the sequential dependence signals accumulated throughout the evolution of the influence path.


\subsubsection{\textbf{Hyperparameter Tuning and Ablation Study}}
We run \textit{IRN} with different combinations of hyperparameter settings and find the optimal group of parameters using grid search. We optimize \textit{IRN} using the Adam algorithm with a dynamic learning rate scheduler which reduces the learning rate by a factor of $2$ once the learning stagnates. 

The ranges and optimal values of the hyperparameters are shown in Table~\ref{tab:hyper}. 
$l_{min}$ and $l_{max}$ refer to the minimum and maximum sequence lengths related to the dataset splitting as introduced in Section \ref{subsec:dataset_split}.
$lr$ denotes the starting learning rate for the model which ranges from $1e-4$ to $1e-2$.
$d$ denotes the embedding size of items, while $d'$ denotes the embedding size of users. $L$ denotes the number of decoder layers, $h$ denotes the number of heads for attention, and $w_t$ is related to the weight of the attention mask for the objective item. 
With the optimal setting of hyperparameters, the training of \textit{IRN} takes approximately 1 hour on Lastfm and 2 hours on Movielens-1M, respectively, using a machine with one GTX-2080Ti GPU. 
In terms of scalability, the computational perplexity of \textit{IRN} is $\mathcal{O}(m^2d + |I|d)$ where $m$ and $I$ refer to the input length and the item set, respectively. 
The framework can be easily paralleled to multiple GPUs to accelerate the computation without significantly affecting the performance.

\begin{table}[tbp]
\renewcommand{\arraystretch}{1.3}

    \caption{Hyperparameter Settings}
    \centering
    \begin{tabular}{llll}
    \toprule
    Name&Range&Lastfm&Movielens-1m\\
    \midrule
    $l_{max}$& [30, 40, 50, 60, 70, 80] & 50 & 60\\
    $l_{min}$& - & 20 & 20\\
    \midrule
    Batch size&\{64, 128, 256, 512\}& 128& 128\\
    $lr$&[1e-4, 1e-2]& 8e-3 & 3e-3\\
    $d$&\{10, 20, 30, 40\}& 40 & 30\\
    $d'$&\{4, 6, 8, 10,12\}& 10 & 10\\
    $L$&\{4, 5, 6, 7, 8\}& 5 & 6\\
    $w_{t}$&\{0, 0.25, 0.5, 0.75, 1\}& 1 & 1\\
    $h$&\{1, 2, 3, 4, 5, 6, 7, 8\}& 4 & 6 \\
    \bottomrule
    \end{tabular}
    \label{tab:hyper}
\end{table}

\subsubsection{\textbf{Case Study}}
In addition to the quantitative studies, we demonstrate the effectiveness of \textit{IRN} with a real case.
We provide a real influence path generated by \textit{IRN} from the Movielens dataset. As is shown in Table \ref{tab:genre_example}, the \textit{Mille bolle blu} with a genre of ``Comedy'' was selected as the objective item while the user's viewing history ends at \textit{Bulletproof} with a genre of ``Action''. In the initial few steps, \textit{IRN} recommended movies in ``Action'', ``Adventure'', and ``Thriller''. After a few swinging attempts between ``Adventure'' and ``Thriller'', the \textit{Space Jam} with genres ``Adventure'', ``Children'', and ``Comedy'' was recommended, followed by \textit{First Kid} with genres ``Children'' and ``Comedy''. The last few movies are classified into ``Comedy''.  From the example, we can observe a smooth genre transition from ``Action'' to ``Comedy''. The results affirm the effectiveness of \textit{IRN} in leading the user towards the objective item smoothly. 

\begin{table}[tb]
\renewcommand{\arraystretch}{1.3}

    \caption{An Example of Movie Genre Shifting (* refers to the objective item). }
    \centering

    \begin{tabular}{ll}
    \toprule
    \bf Name & \bf Genre\\
    \midrule
    \multicolumn{2}{l}{\bf The last movie in the viewing history:}\\
    Bulletproof (1996) & Action\\
    \midrule
    \multicolumn{2}{l}{\bf The movies in the influence path:}\\
    Lost World: Jurassic Park (1997)	&		Action, Adventure, Sci-Fi, Thriller\\
    Chain Reaction (1996)	&		Action, Adventure, Thriller\\
    Anaconda (1997)		&	Action, Adventure, Thriller\\
    Broken Arrow (1996)	&		Action, Thriller\\
    Fled (1996)	&		Action, Adventure\\
    Glimmer Man (1996)	&		Action, Thriller\\
    Space Jam (1996)	&		Adventure, Animation, Children,\\ 
                        &       Comedy, Fantasy\\
    First Kid (1996)	&		Children, Comedy\\
    Jack (1996)		&	Comedy, Drama\\
    Michael (1996)	&		Comedy, Romance\\
    \textbf{Mille bolle blu (1993)}*	&		Comedy\\
    \bottomrule
    \end{tabular}
    \label{tab:genre_example}
\end{table}

\section{Conclusion}
\label{sec:conclusion}
We introduce a new recommendation paradigm termed \textit{Influential Recommender System} (\textit{IRS}), which proactively and incrementally expands a user's interest toward an objective item by recommending to the user a sequence of items. We design a Transformer-based sequential model \textit{Influential Recommender Network} (\textit{IRN}) with \textit{Personalized Impressionability Mask} (\textit{PIM}) to model, respectively, the sequential dependencies among the items and individual users' personalized acceptance of external influence. Evaluation metrics are carefully designed for the influential recommendation task. Experimental results show that \textit{IRN} significantly outperforms the baseline recommenders and demonstrate \textit{IRN}'s capability in smooth and progressive persuasion.

Since this is the first work that proposes the idea of influential recommendation, we are in the exploratory stage for this new recommendation problem. Many issues remain to be further studied.
We summarize the future work in the following directions.
(1) Extend the path-finding baseline by incorporating knowledge graphs (\textit{KG}). Knowledge graphs have been extensively used in recommendation systems, making them a natural fit of the \textit{IRS} framework. For example, we can model the user's historical interests as a subgraph and expand the subgraph toward the objective item.
(2) Improve the evaluation of unknown sequence-item interactions. Due to the limitation of the offline dataset used in this work, we utilize an evaluator to stimulate the user's interest in a new incoming item, which may incur bias in calculating the \textit{IRS} evaluation metrics. A better solution is to deploy the \textit{IRS} in an online recommendation platform where honest user feedback is available or to conduct a human experiment to evaluate the generated influence paths.
(3) Expand the scope of the objective in \textit{IRS}. In this work, the proposed framework aims at leading the user to like a single objective item, which naturally leads to a broader problem setting where the objective can be a collection of items, a category, a topic, etc. 
(4) Consider the stepwise dynamics in generating the influence path. For simplicity, we assume that the user passively accepts the recommended items and evaluates the influence path as a whole. However, in a more practical application scenario, it is important to consider the stepwise user response to a recommended item in approaching the objective. For example, a user may reject an intermediate item in a pre-scheduled influence path, and the \textit{IRS} needs to alter its strategy by recommending another item to persuade the user towards the objective.





 \section*{Acknowledgment}
 The authors would thank anonymous reviewers for their valuable comments and suggestions. This work was sponsored by Research Grants Council HKSAR GRF (No. 16215019), NSFC (No. 62102244), CCF-Tencent Open Research Fund (RAGR20220129), and Guang Dong Innovative Yong Talents Program (No. R5201919).

\bibliographystyle{abbrv}
\bibliography{citation} 

\begin{thebibliography}{10}

\bibitem{ajzen2012attitudes}
I.~Ajzen.
\newblock Attitudes and persuasion.
\newblock 2012.

\bibitem{arieli2019private}
I.~Arieli and Y.~Babichenko.
\newblock Private bayesian persuasion.
\newblock {\em Journal of Economic Theory}, 182:185--217, 2019.

\bibitem{barkan2016item2vec}
O.~Barkan and N.~Koenigstein.
\newblock Item2vec: neural item embedding for collaborative filtering.
\newblock In {\em 2016 IEEE 26th International Workshop on Machine Learning for
  Signal Processing (MLSP)}, pages 1--6. IEEE, 2016.

\bibitem{bergemann2016information}
D.~Bergemann and S.~Morris.
\newblock Information design, bayesian persuasion, and bayes correlated
  equilibrium.
\newblock {\em American Economic Review}, 106(5):586--91, 2016.

\bibitem{bharadhwaj2018recgan}
H.~Bharadhwaj, H.~Park, and B.~Y. Lim.
\newblock Recgan: recurrent generative adversarial networks for recommendation
  systems.
\newblock In {\em Proceedings of the 12th ACM Conference on Recommender
  Systems}, pages 372--376, 2018.

\bibitem{chaiken1987heuristic}
S.~Chaiken.
\newblock The heuristic model of persuasion.
\newblock In {\em Social influence: the ontario symposium}, volume~5, pages
  3--39, 1987.

\bibitem{che2018recommender}
Y.-K. Che and J.~H{\"o}rner.
\newblock Recommender systems as mechanisms for social learning.
\newblock {\em The Quarterly Journal of Economics}, 133(2):871--925, 2018.

\bibitem{chen2019top}
M.~Chen, A.~Beutel, P.~Covington, S.~Jain, F.~Belletti, and E.~H. Chi.
\newblock Top-k off-policy correction for a reinforce recommender system.
\newblock In {\em Proceedings of the Twelfth ACM International Conference on
  Web Search and Data Mining}, pages 456--464, 2019.

\bibitem{dwarampudi2019effects}
M.~Dwarampudi and N.~Reddy.
\newblock Effects of padding on lstms and cnns.
\newblock {\em arXiv preprint arXiv:1903.07288}, 2019.

\bibitem{he2017translation}
R.~He, W.-C. Kang, and J.~McAuley.
\newblock Translation-based recommendation.
\newblock In {\em Proceedings of the eleventh ACM conference on recommender
  systems}, pages 161--169, 2017.

\bibitem{he2016fusing}
R.~He and J.~McAuley.
\newblock Fusing similarity models with markov chains for sparse sequential
  recommendation.
\newblock In {\em 2016 IEEE 16th International Conference on Data Mining
  (ICDM)}, pages 191--200. IEEE, 2016.

\bibitem{he2017neural}
X.~He, L.~Liao, H.~Zhang, L.~Nie, X.~Hu, and T.-S. Chua.
\newblock Neural collaborative filtering.
\newblock In {\em Proceedings of the 26th international conference on world
  wide web}, pages 173--182, 2017.

\bibitem{hidasi2018recurrent}
B.~Hidasi and A.~Karatzoglou.
\newblock Recurrent neural networks with top-k gains for session-based
  recommendations.
\newblock In {\em Proceedings of the 27th ACM international conference on
  information and knowledge management}, pages 843--852, 2018.

\bibitem{hidasi2015session}
B.~Hidasi, A.~Karatzoglou, L.~Baltrunas, and D.~Tikk.
\newblock Session-based recommendations with recurrent neural networks.
\newblock {\em arXiv preprint arXiv:1511.06939}, 2015.

\bibitem{Huang_simulator2020}
J.~Huang, H.~Oosterhuis, M.~de~Rijke, and H.~van Hoof.
\newblock {\em Keeping Dataset Biases out of the Simulation: A Debiased
  Simulator for Reinforcement Learning Based Recommender Systems}, page
  190–199.
\newblock 2020.

\bibitem{huang2020keeping}
J.~Huang, H.~Oosterhuis, M.~de~Rijke, and H.~van Hoof.
\newblock Keeping dataset biases out of the simulation: A debiased simulator
  for reinforcement learning based recommender systems.
\newblock In {\em Fourteenth ACM Conference on Recommender Systems}, pages
  190--199, 2020.

\bibitem{jadbabaie2012non}
A.~Jadbabaie, P.~Molavi, A.~Sandroni, and A.~Tahbaz-Salehi.
\newblock Non-bayesian social learning.
\newblock {\em Games and Economic Behavior}, 76(1):210--225, 2012.

\bibitem{kamenica2019bayesian}
E.~Kamenica.
\newblock Bayesian persuasion and information design.
\newblock {\em Annual Review of Economics}, 11:249--272, 2019.

\bibitem{kamenica2011bayesian}
E.~Kamenica and M.~Gentzkow.
\newblock Bayesian persuasion.
\newblock {\em American Economic Review}, 101(6):2590--2615, 2011.

\bibitem{kang2018self}
W.-C. Kang and J.~McAuley.
\newblock Self-attentive sequential recommendation.
\newblock In {\em 2018 IEEE International Conference on Data Mining (ICDM)},
  pages 197--206. IEEE, 2018.

\bibitem{li2021easy}
G.~Li, Y.~Xi, J.~Ding, D.~Wang, B.~Liu, C.~Fan, X.~Mao, and Z.~Zhao.
\newblock Easy and efficient transformer: Scalable inference solution for large
  nlp model.
\newblock {\em arXiv preprint arXiv:2104.12470}, 2021.

\bibitem{li2018multi}
J.~Li, Z.~Tu, B.~Yang, M.~R. Lyu, and T.~Zhang.
\newblock Multi-head attention with disagreement regularization.
\newblock {\em arXiv preprint arXiv:1810.10183}, 2018.

\bibitem{li2015counterfactual}
L.~Li, S.~Chen, J.~Kleban, and A.~Gupta.
\newblock Counterfactual estimation and optimization of click metrics in search
  engines: A case study.
\newblock In {\em Proceedings of the 24th International Conference on World
  Wide Web}, pages 929--934, 2015.

\bibitem{qiu2020pre}
X.~Qiu, T.~Sun, Y.~Xu, Y.~Shao, N.~Dai, and X.~Huang.
\newblock Pre-trained models for natural language processing: A survey.
\newblock {\em Science China Technological Sciences}, pages 1--26, 2020.

\bibitem{rendle2012bpr}
S.~Rendle, C.~Freudenthaler, Z.~Gantner, and L.~Schmidt-Thieme.
\newblock Bpr: Bayesian personalized ranking from implicit feedback.
\newblock {\em arXiv preprint arXiv:1205.2618}, 2012.

\bibitem{rendle2010factorizing}
S.~Rendle, C.~Freudenthaler, and L.~Schmidt-Thieme.
\newblock Factorizing personalized markov chains for next-basket
  recommendation.
\newblock In {\em Proceedings of the 19th international conference on World
  wide web}, pages 811--820, 2010.

\bibitem{romanyuk2019cream}
G.~Romanyuk and A.~Smolin.
\newblock Cream skimming and information design in matching markets.
\newblock {\em American Economic Journal: Microeconomics}, 11(2):250--76, 2019.

\bibitem{sun2019bert4rec}
F.~Sun, J.~Liu, J.~Wu, C.~Pei, X.~Lin, W.~Ou, and P.~Jiang.
\newblock Bert4rec: Sequential recommendation with bidirectional encoder
  representations from transformer.
\newblock In {\em Proceedings of the 28th ACM international conference on
  information and knowledge management}, pages 1441--1450, 2019.

\bibitem{Tang2019}
J.~Tang, F.~Belletti, S.~Jain, M.~Chen, A.~Beutel, C.~Xu, and E.~H.~Chi.
\newblock Towards neural mixture recommender for long range dependent user
  sequences.
\newblock {\em The World Wide Web Conference on - WWW ’19}, 2019.

\bibitem{tang2018personalized}
J.~Tang and K.~Wang.
\newblock Personalized top-n sequential recommendation via convolutional
  sequence embedding, 2018.

\bibitem{vaswani2017attention}
A.~Vaswani, N.~Shazeer, N.~Parmar, J.~Uszkoreit, L.~Jones, A.~N. Gomez,
  {\L}.~Kaiser, and I.~Polosukhin.
\newblock Attention is all you need.
\newblock In {\em Advances in neural information processing systems}, pages
  5998--6008, 2017.

\bibitem{wang2019sequential}
F.~Wang, X.~Fang, L.~Liu, Y.~Chen, J.~Tao, Z.~Peng, C.~Jin, and H.~Tian.
\newblock Sequential evaluation and generation framework for combinatorial
  recommender system.
\newblock {\em arXiv preprint arXiv:1902.00245}, 2019.

\bibitem{wang2018billion}
J.~Wang, P.~Huang, H.~Zhao, Z.~Zhang, B.~Zhao, and D.~L. Lee.
\newblock Billion-scale commodity embedding for e-commerce recommendation in
  alibaba.
\newblock In {\em Proceedings of the 24th ACM SIGKDD International Conference
  on Knowledge Discovery \& Data Mining}, pages 839--848, 2018.

\bibitem{wang2018attentionbased}
S.~Wang, L.~Hu, L.~Cao, X.~Huang, D.~Lian, and W.~Liu.
\newblock Attention-based transactional context embedding for next-item
  recommendation.
\newblock 2018.

\bibitem{ShoujingWang-ijcai2019}
S.~Wang, L.~Hu, Y.~Wang, L.~Cao, Q.~Z. Sheng, and M.~Orgun.
\newblock Sequential recommender systems: Challenges, progress and prospects.
\newblock In {\em Proceedings of the Twenty-Eighth International Joint
  Conference on Artificial Intelligence, {IJCAI-19}}, pages 6332--6338, 7 2019.

\bibitem{Wang2019}
S.~Wang, L.~Hu, Y.~Wang, Q.~Z. Sheng, M.~Orgun, and L.~Cao.
\newblock Modeling multi-purpose sessions for next-item recommendations via
  mixture-channel purpose routing networks.
\newblock In {\em Proceedings of the Twenty-Eighth International Joint
  Conference on Artificial Intelligence, {IJCAI-19}}, pages 3771--3777.
  International Joint Conferences on Artificial Intelligence Organization, 7
  2019.

\bibitem{wang2020global}
Z.~Wang, W.~Wei, G.~Cong, X.-L. Li, X.-L. Mao, and M.~Qiu.
\newblock Global context enhanced graph neural networks for session-based
  recommendation.
\newblock In {\em Proceedings of the 43rd International ACM SIGIR Conference on
  Research and Development in Information Retrieval}, pages 169--178, 2020.

\bibitem{Wu2017RRN}
C.-Y. Wu, A.~Ahmed, A.~Beutel, A.~J. Smola, and H.~Jing.
\newblock Recurrent recommender networks.
\newblock WSDM '17, pages 495--503. ACM, 2017.

\bibitem{wu2019session}
S.~Wu, Y.~Tang, Y.~Zhu, L.~Wang, X.~Xie, and T.~Tan.
\newblock Session-based recommendation with graph neural networks.
\newblock In {\em Proceedings of the AAAI conference on artificial
  intelligence}, volume~33, pages 346--353, 2019.

\bibitem{wu2016collaborative}
Y.~Wu, C.~DuBois, A.~X. Zheng, and M.~Ester.
\newblock Collaborative denoising auto-encoders for top-n recommender systems.
\newblock In {\em Proceedings of the Ninth ACM International Conference on Web
  Search and Data Mining}, pages 153--162, 2016.

\bibitem{xiang2010temporal}
L.~Xiang, Q.~Yuan, S.~Zhao, L.~Chen, X.~Zhang, Q.~Yang, and J.~Sun.
\newblock Temporal recommendation on graphs via long-and short-term preference
  fusion.
\newblock In {\em Proceedings of the 16th ACM SIGKDD international conference
  on Knowledge discovery and data mining}, pages 723--732, 2010.

\bibitem{Ying2018}
H.~Ying, F.~Zhuang, F.~Zhang, Y.~Liu, G.~Xu, X.~Xie, H.~Xiong, and J.~Wu.
\newblock Sequential recommender system based on hierarchical attention
  network.
\newblock In {\em Proceedings of the 27th International Joint Conference on
  Artificial Intelligence}, IJCAI'18, page 3926–3932. AAAI Press, 2018.

\bibitem{Yuan2019}
F.~Yuan, A.~Karatzoglou, I.~Arapakis, J.~M. Jose, and X.~He.
\newblock A simple convolutional generative network for next item
  recommendation.
\newblock WSDM '19, page 582–590, 2019.

\bibitem{yue2011linear}
Y.~Yue and C.~Guestrin.
\newblock Linear submodular bandits and their application to diversified
  retrieval.
\newblock Neural Information Processing Systems, 2011.

\bibitem{zhang2019feature}
T.~Zhang, P.~Zhao, Y.~Liu, V.~S. Sheng, J.~Xu, D.~Wang, G.~Liu, and X.~Zhou.
\newblock Feature-level deeper self-attention network for sequential
  recommendation.
\newblock In {\em IJCAI}, pages 4320--4326, 2019.

\bibitem{zhao2016much}
P.~Zhao and D.~L. Lee.
\newblock How much novelty is relevant? it depends on your curiosity.
\newblock In {\em Proceedings of the 39th International ACM SIGIR conference on
  Research and Development in Information Retrieval}, pages 315--324, 2016.

\end{thebibliography}

\end{document}